\documentclass[%
 preprint,
 superscriptaddress,
 amsmath,amssymb,
 aps,
 showkeys,
 titlepage,
 endfloats*.   
]{revtex4-2}
\usepackage[utf8]{inputenc}
\usepackage[english]{babel}
\usepackage{amsmath}
\usepackage{graphicx}
\usepackage{dcolumn}
\usepackage{bm}

\usepackage[colorlinks = true,
            linkcolor = blue,
            urlcolor  = blue,
            citecolor = red,
            anchorcolor = blue]{hyperref}

\begin{document}


\title{Anharmonic Quantum Transport Analysis of Thermal Transport Anomalies in Ultrathin Silicon Nanowires} 


\author{Lokanath Patra}
\affiliation{%
George W. Woodruff School of Mechanical Engineering, Georgia Institute of Technology, Atlanta, GA 30332, USA}%

\author{Mayur Pratap Singh}
\affiliation{%
George W. Woodruff School of Mechanical Engineering, Georgia Institute of Technology, Atlanta, GA 30332, USA}%

\author{Satish Kumar}
\email{satish.kumar@me.gatech.edu} 
\affiliation{%
George W. Woodruff School of Mechanical Engineering, Georgia Institute of Technology, Atlanta, GA 30332, USA}%

\date{\today}

\begin{abstract}
Thermal transport in low-dimensional semiconductors is crucial for advancing thermal management in nanoelectronics, quantum devices, and thermoelectric devices. Recent molecular dynamics (MD) studies have identified a nonmonotonic dependence of thermal conductivity ($\kappa$) on diameter in ultrathin silicon nanowires (NWs). However, classical MD methods are limited at low temperatures and in strongly confined regimes. This work introduces a fully quantum-mechanical perspective on this anomaly by employing anharmonic non-equilibrium Green's function (NEGF) simulations combined with density functional theory-trained neuroevolution potentials. All reported transport results correspond to NW devices with a finite channel length of \(L=10\) nm and are therefore expressed as effective thermal conductivities (\(\kappa_{\mathrm{eff}})\). At 300 K, \(\kappa_{\mathrm{eff}}\) decreases with diameter to minima at \(d_c^{300\mathrm{K}}\approx 6.24\) nm and 5.50 nm for [001]- and [110]-oriented NWs, respectively. At 10 K, the corresponding minima occur at \(d_c^{10\mathrm{K}}\approx 7.33\) nm and 5.50 nm. At room temperature, this behavior arises from dominant momentum-conserving normal scattering relative to Umklapp processes in confined regimes, thereby enabling Poiseuille-like hydrodynamic phonon flow that competes with boundary scattering. At cryogenic temperatures, Bose–Einstein statistics restrict heat transport to low-frequency confined acoustic-like modes. Comparison with the harmonic Landauer limit shows that anharmonic phonon–phonon scattering provides a finite correction to \(\kappa_{\mathrm{eff}}\). In contrast to classical MD, which is inaccurate at low temperatures due to overexcitation of high-frequency vibrations by Boltzmann statistics, neglect of quantum suppression, and overestimation of thermal conductivity in thinner NWs with stronger quantum confinement, the NEGF framework provides quantitative accuracy even at low temperatures, such as 10 K.
\end{abstract}

                            
\maketitle


\section{Introduction}
Materials at the micro- and nanoscale exhibit distinctive thermal transport properties determined by the interplay of phonon scattering, interface interactions, and quantum confinement effects~\cite{cahill2003nanoscale,nomura2022review,zhang2018theoretical,beardo2025nanoscale,machida2020phonon,siemens2010quasi}. These phenomena pose significant fundamental challenges and simultaneously create opportunities for applications in integrated circuits ~\cite{shakouri2006nanoscale}, thermoelectric energy conversion~\cite{hochbaum2008enhanced,chen2019thermoelectrics}, and sensing technology~\cite{chang2023nanowire}. Among low-dimensional systems, semiconductor nanowires (NWs) serve as model structures for investigating size-dependent thermal transport due to their well-defined geometry and tunable physical properties~\cite{hochbaum2010semiconductor,swinkels2018nanowires,dasgupta201425th}. Experimental studies have shown that Si NWs can achieve a thermoelectric figure of merit nearly two orders of magnitude greater than that of bulk silicon~\cite{hochbaum2008enhanced,boukai2008silicon}. This enhancement is primarily due to strong phonon-boundary scattering, which substantially reduces thermal conductivity ($\kappa$) in the NW form. 

However, recent molecular dynamics (MD) studies reveal conflicting trends in the dependence of NW diameter on $\kappa$ in the ultrathin regime~\cite{zhou2017nonmonotonic,ponomareva2007thermal,donadio2010temperature,rashid2018effect,xu2025critical,zaccone2025phonon}. For example, using Green-Kubo equilibrium MD simulations, Zhou et al.~\cite{zhou2017nonmonotonic} reported that below a critical diameter ($d_c$) of 2-3 nm, the $\kappa$ increases with decreasing diameter for ultrathin Si NWs. Remarkably, $\kappa$ increases dramatically for the thinnest thermodynamically stable Si NWs (diameters below 1 nm), reaching up to 13 times that of bulk silicon. Using the equilibrium Green–Kubo approach, Donadio and Galli also reported that Si NWs with a diameter of 1.1 nm have higher thermal conductivity than those with diameters of 2 and 3 nm~\cite{donadio2010temperature}. By combining atomistic simulations with the relaxation-time approximation to the phonon Boltzmann transport equation, Rashid \textit{et al.}~\cite{rashid2018effect} found that the lattice thermal conductivity in ultrathin Si NWs is substantially lower than that in bulk Si, with no nonmonotonic behavior observed. In a recent MD-based study, Xu \textit{et al.}~\cite{xu2025critical} found a nonmonotonic dependence of $\kappa$ on Si NW diameter using a machine-learning potential (MLP); they also observed a complete structural transformation to an amorphous phase for diameters below 1.1 nm. Theoretical evidence for this nonmonotonic behavior has been reported in other low-dimensional systems, including diamond~\cite{liang2022abnormally} and GaP NWs~\cite{gireesan2020diameter}, highlighting the fundamental role of phonon hydrodynamics in nanoscale thermal transport. Similar nonmonotonic trends in $\kappa$ with thickness have also been observed in layered two-dimensional materials such as MoS$_2$~\cite{yuan2018nonmonotonic}, TiS$_3$~\cite{liu2022non}, and WTe$_2$~\cite{wu2020anomalous}.

The complex dynamics of nanoscale thermal transport present significant challenges. MD simulations are widely utilized and often represent the only practical method for investigating thermal transport in NW systems~\cite{dong2024molecular,wang2007quantum}. However, MD simulations are inherently limited, as they cannot capture quantum-level thermal transport phenomena.~\cite{zeng2018nanoscale} Moreover, inaccuracies in empirical interatomic potentials limit their ability to capture the effects of surface modifications, disorder formation, and phonon hydrodynamics on heat transport in Si NWs, often leading to unphysical or inconsistent results. At very low temperatures, where quantum effects predominate, these transport regimes remain inaccessible to any MD simulation, regardless of the interatomic potential applied. Therefore, alternative approaches that are both accurate and efficient, such as quantum transport theory, are necessary. Among these, the NEGF formalism has emerged as a robust tool for atomistic, quantum-mechanical analysis of heat transport.~\cite{datta2005quantum,chen2018thermal,guo2020quantum,guo2021anharmonic,yan2016role,mingo2006anharmonic}

In this study, we have performed fully anharmonic NEGF~\citep{guo2020quantum,guo2021anharmonic,luisier2012atomistic} simulations for [001] and [110] Si NWs in which both harmonic and third-order anharmonic force constants are obtained from a machine-learning neural-network potential (NEP)~\cite{fan2021neuroevolution,fan2022gpumd,fan2022improving,singh2026revealing} that was explicitly trained on high-accuracy DFT reference data (Fig.~S1 in supplementary information (SI)). To our knowledge, this is the first work to integrate NEP with anharmonic NEGF simulations. This hybrid DFT$\to$NEP$\to$NEGF workflow reduces the computational cost by several orders of magnitude, enabling systematic, high-throughput thermal transport calculations across a wide range of diameters (0.5--6 nm) at both 300 K and 10 K. We reproduce the counterintuitive rise of effective thermal conductivity \(\kappa_{\mathrm{eff}}\) values with decreasing $d$ in the ultrathin limit, revealing a nonmonotonic \(\kappa_{\mathrm{eff}}(d)\) dependence at both 300 K and 10 K. The critical diameters are denoted by \(d_c^{300\mathrm{K}}\) and \(d_c^{10\mathrm{K}}\) for the room-temperature and cryogenic results, respectively. At room temperature, this anomaly arises from the competition between boundary scattering, dominant momentum-conserving normal (N) processes, and resistive Umklapp (U) scattering, facilitating Poiseuille-like phonon hydrodynamic flow in the confined regime. At cryogenic temperatures, radial quantum confinement discretizes the low-frequency phonon spectrum, while Bose–Einstein statistics restrict heat transport to a small set of confined acoustic-like modes. Comparison with harmonic Landauer calculations further quantifies the finite anharmonic correction in this regime. The NEGF formalism ensures quantitative accuracy even at these cryogenic conditions, which are inaccessible to MD.

The fully anharmonic NEGF method employed here offers distinct advantages over classical MD approaches, particularly in the ultrathin regime and at low temperatures. MD simulations often yield high $\kappa$ values because they neglect quantum statistics, and empirical force fields also struggle to accurately describe highly confined phonon dispersions and surface-modified scattering in such small structures. In contrast, the quantum-mechanical NEGF formalism, combined with DFT-trained NEP force constants, provides a rigorous treatment of phonon transmission, full quantum statistics, and higher-order anharmonicity. This enables quantitatively reliable results across the entire diameter range (down to 0.5 nm) and at cryogenic temperatures (e.g., 10 K), where classical MD is inapplicable because it cannot describe quantum-coherent transport regimes.

\section{Results and Discussions}
\begin{figure}[!t]
    \centering
    \includegraphics[width=\linewidth]{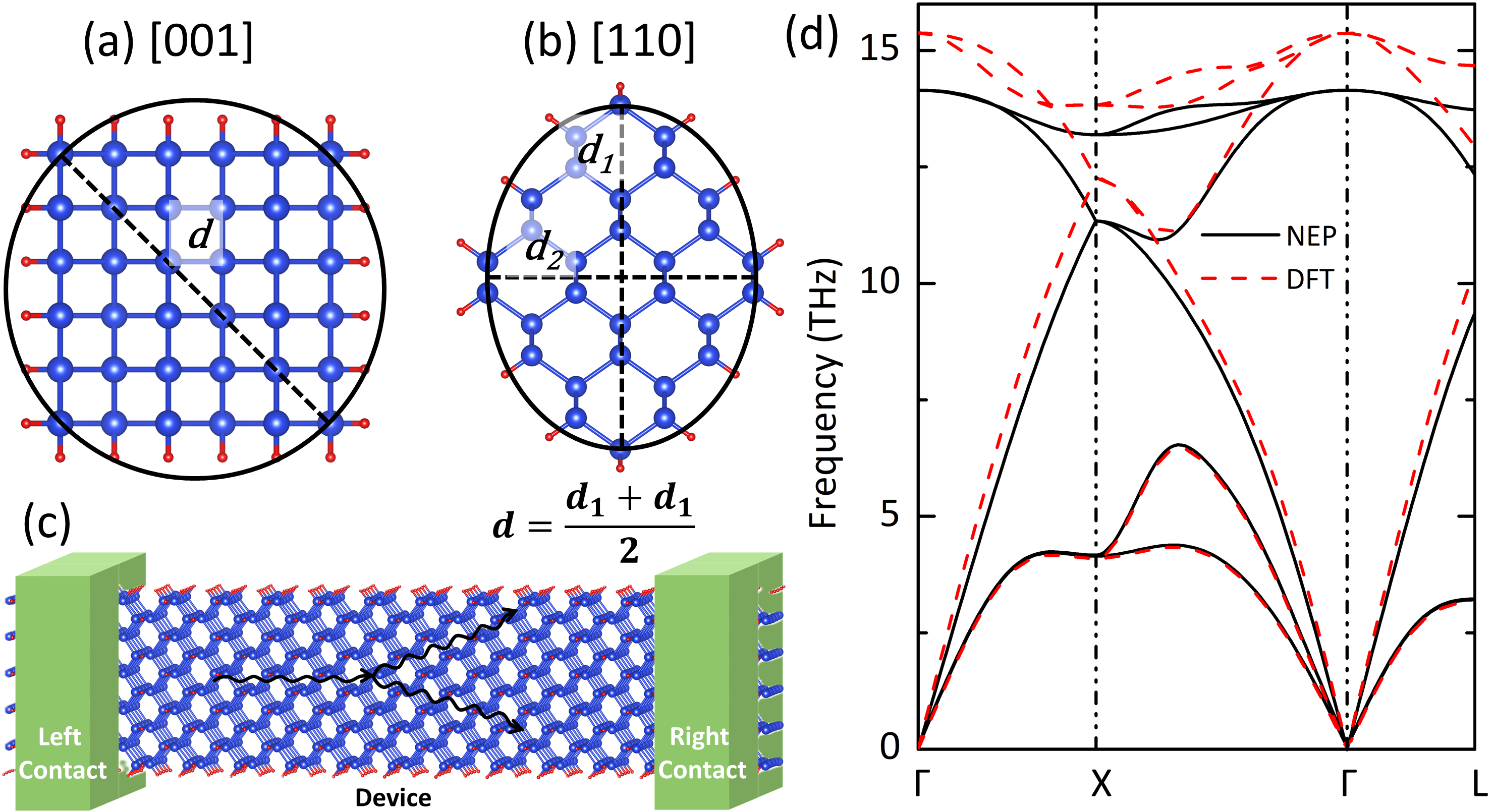}
    \caption{Top and side views of the atomic structure of (a) [001]- and (b) [110]-oriented silicon NWs. The NW surfaces are passivated with hydrogen. For [001] NWs, the diameter ($d$) is defined as the corner-to-corner distance of the rectangular cross-section; for [110] NWs, the diameter ($d$) is taken as the average of the major ($d_1$) and minor ($d_2$) axes of the approximately oval cross-section. (c)  Schematic of the model device used in the NEGF simulations. The central channel consists of a silicon NW, with semi-infinite left and right contacts (green regions). (d) Phonon band structure of FCC silicon calculated using DFT and the trained NEP.}
    \label{fig:device}
\end{figure}

Atomic structures of ultrathin Si NWs are generated from bulk face-centered cubic (FCC) Si with an optimized lattice parameter of 5.44 \AA, using QuantumATK software~\cite{smidstrup2019quantumatk}. The SI provides the detailed theory and computational techniques. The NWs are modeled along [001] and [110] crystallographic directions, which are most frequently studied experimentally~\cite{li2003thermal,lee2016thermal} and theoretically~\cite{zhou2017nonmonotonic,xu2025critical,hahn2023thermal,he2012microscopic}. Figures~\ref{fig:device}(a) and (b) show the top and side views of [001] and [110] Si NWs, respectively, with surface passivation achieved using hydrogen atoms. Passivating with hydrogen atoms is a common approach to stabilizing nanostructures by reducing surface reconstruction and dangling bonds.~\cite{gao2022catalyst} Since the cross-sectional shapes are not perfectly circular, the effective diameter $d$ is defined as the corner-to-corner distance for [001]-oriented NWs and as the average of the major axis ($d_1$) and minor axis ($d_2$) for the approximately elliptical [110]-oriented NWs.The length of all NWs is fixed at 10 nm, regardless of diameter; therefore, the reported conductivities are effective finite-length values derived from the calculated two-terminal conductance as \(\kappa_{\mathrm{eff}}=GL/A\). Here, \(G\) denotes the two-terminal thermal conductance, and \(A\) denotes the effective cross-sectional area. Throughout this work, we refer to NWs by their corresponding diameter \textit{d}. Figure~\ref{fig:device}(c) presents the device model employed in our anharmonic NEGF simulations. To address the computational challenges of third-order force-constant calculations, we have developed an ML-based Si-H NEP. The NEP reproduces the overall DFT phonon dispersion, including the low-frequency acoustic branches most relevant to cryogenic transport, while visible deviations remain in portions of the optical branches [Fig.~\ref{fig:device}(d)]. Such deviations are commonly reported for machine-learning interatomic potentials trained on finite configuration datasets, particularly for high-frequency optical modes.~\cite{xu2025critical,yin2026accelerated,wu2026atomic,cheng2025revealing} Details of the training procedure are provided in the SI. We then used the obtained force constants to study phonon-phonon scattering using our in-house anharmonic NEGF framework. These ML-NEGF simulations serve as an efficient alternative to conventional DFT-NEGF approaches, reducing computational cost by 100$\times$.  

\begin{figure}[!tb]
    \centering
    \includegraphics[width=1\linewidth]{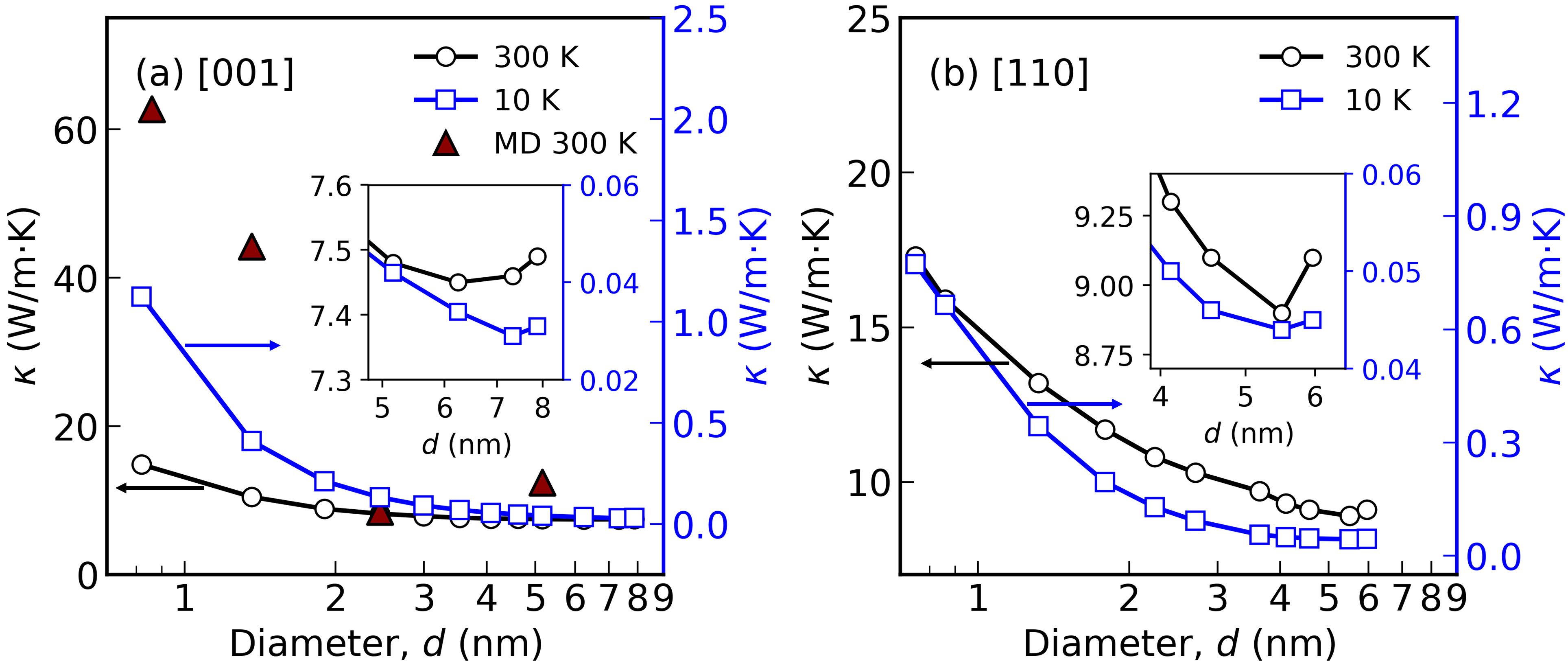}
    \caption{Diameter ($d$) dependence of the effective finite-length thermal conductivity ($\kappa_{eff}=GL/A$) of (a) [001]- and (b) [110]-oriented Si NWs at 300 K (black circles, left axis) and 10 K (blue squares, right axis), computed using the ML-NEGF approach. The insets zoom into the large-diameter regime (highlighted by the red rectangles in the main panels) and identify the temperature-specific critical diameters. At 300 K, \(d_c^{300\mathrm{K}}\approx 6.24\) nm for [001] and \(\approx 5.50\) nm for [110]. At 10 K, \(d_c^{10\mathrm{K}}\approx 7.33\) nm for [001] and \(\approx 5.50\) nm for [110]. For comparison, MD results along the [001] direction are shown as red triangles. All calculations use a channel length of \(L=10\) nm; the length dependence for a representative NW is provided in Fig. S3.}
    \label{fig:conductivity}
\end{figure}

The effective thermal conductivities, \(\kappa_{\mathrm{eff}}\), as a function of $d$ at 10 K and 300 K for NWs along [001] and [110] directions are shown in Fig.~\ref{fig:conductivity}(a) and (b), respectively. Consistent with previous studies on Si and GaP NWs~\cite{zhou2017nonmonotonic,xu2025critical,gireesan2020diameter}, we observe a nonmonotonic dependence of $\kappa$ on $d$. For the thinnest considered $d$, our methods estimate $\kappa$ to be 14.9 $\mathrm{W/m\!\cdot\!K}$ and 17.3 $\mathrm{W/m\!\cdot\!K}$ at 300 K for [001] and [110] NWs, respectively. The calculated $\kappa$ for [001] NWs is significantly lower, by 2 to 3 orders of magnitude, compared to the values reported by Zhou \textit{et al}.~\cite{zhou2017nonmonotonic} for similar diameters. The discrepancy is primarily due to the different interatomic potentials used in their study, including the Tersoff environment-dependent potential and the Stillinger–Weber potential. However, in a recent MD-based study, Xu \textit{et al.} estimated a $\kappa$ value of $\sim 9~\mathrm{W/m\!\cdot\!K}$ using an interatomic NEP,~\cite{xu2025critical} which is close to our calculated value. Additionally, for larger diameters, the estimated $\kappa$ values agree well with both experimental data~\cite{li2003thermal} and theoretical predictions by Dames and Chen~\cite{dames2004theoretical}, further validating the accuracy of the trained NEP for studying the thermal properties of Si-based materials. For instance, at 300\,K, the lowest measured diameter of 22\,nm gives $\kappa \sim 8~\mathrm{W/m\!\cdot\!K}$, while the theoretical prediction yields $10\ \mathrm{W/m\!\cdot\!K}$ for a 10\,nm diameter. Our calculation gives $\kappa \approx 7.5\ \mathrm{W/m\!\cdot\!K}$ for a 7.9\,nm diameter, indicating a scale comparable to both experiment and theory. Similarly, at T = 10 K, $\kappa(d)$ exhibits nonmonotonic behavior for both crystallographic orientations (Fig.~\ref{fig:conductivity}). Unlike classical MD simulations, which become unreliable at low temperatures, the present NEGF framework fully accounts for quantum-mechanical phonon effects, enabling accurate predictions of $\kappa$ down to cryogenic temperatures.

The length dependence of \(\kappa_{\mathrm{eff}}\) was assessed for a [001]-oriented NW with \(d=6.24\) nm. As shown in Fig. S3, \(\kappa_{\mathrm{eff}}\) increases as the channel length is increased from 10 to 20 nm, demonstrating that transport remains length-dependent over the investigated range. A systematic length-convergence study for all diameters is computationally prohibitive within the fully anharmonic NEGF framework. Figure S4 evaluates the numerical uncertainty of the harmonic NEGF method by systematically varying the \(k\)-point sampling and the \(\sigma \) parameter for representative \([001]\)-oriented NWs.

At 300 K, as shown in Fig. 2, the calculated \(\kappa_{\mathrm{eff}}\) of [001]- and [110]-oriented Si NWs exhibits a nonmonotonic dependence on diameter \(d\): \(\kappa_{\mathrm{eff}}\) decreases with increasing \(d\), reaches a minimum at \(d_c^{300\mathrm{K}}\approx 6.24\) nm for [001] and \(\approx 5.50\) nm for [110], and increases thereafter. These $d_c$ values are higher than those typically reported in classical MD studies ($\sim$2--3\,nm, depending on the interatomic potential).~\cite{zhou2017nonmonotonic,xu2025critical} To further validate our findings, we performed MD simulations on [001]-oriented Si NWs using the developed NEP. The resulting thermal conductivities for selected representative diameters are plotted alongside the NEGF data in Fig.~\ref{fig:conductivity}(a). For diameters $d > 2\,\text{nm}$, the MD results align closely with the NEGF calculations. However, in the ultrathin limit ($d \lesssim 2\,\text{nm}$), MD yields $\kappa$ values that are significantly higher than those from NEGF. This discrepancy arises because classical MD follows Boltzmann statistics, fully excites high-frequency modes even at moderate temperatures, and neglects quantum suppression via Bose-Einstein occupation. As a result, classical MD typically shows nonmonotonic $\kappa$ vs. $d$ behavior only in extremely thin NWs. In that ultrathin regime, classical MD overestimates thermal transport due to artificial effects, such as enhanced hydrodynamic phonon flow (driven by strong normal scattering) competing with surface scattering.~\cite{zhou2017nonmonotonic} In contrast, anharmonic NEGF accurately incorporates quantum statistics, frequency-dependent phonon transmission, and confinement-enhanced Umklapp scattering. This yields a more realistic description, shifting the minimum to larger $d_c$, where additional low-frequency subbands that emerge with increasing diameter face stronger anharmonic limitations.

The nonmonotonic behavior of $\kappa(d)$ is unusual compared to the typical monotonic decrease in $\kappa$ with decreasing $d$ observed in most nanostructures, which is primarily attributed to enhanced boundary scattering. The present results highlight the competing roles of boundary scattering, N-scattering, and U-scattering, as captured by Matthiessen's rule for the total phonon relaxation time $\tau$:
\begin{equation}
    \frac{1}{\tau} = \frac{1}{\tau_{\text{bound}}} + \frac{1}{\tau_{N}} + \frac{1}{\tau_{U}},
\label{M-rule}
\end{equation}
where $\tau_{\text{bound}} \approx d / v_g$ in the Casimir limit for diffuse boundary scattering~\cite{casimir1938note} ($v_g$ is the average group velocity), and $\tau_{N}$ and $\tau_{U}$ are the normal and Umklapp relaxation times, respectively. Since $\kappa \propto \int C(\omega) v_g^2(\omega) \tau(\omega) \, d\omega$, the diameter dependence of the scattering rate $\gamma = 1/\tau$ directly governs the observed trend in $\kappa(d)$.

\begin{figure}
    \centering
    \includegraphics[width=1\linewidth]{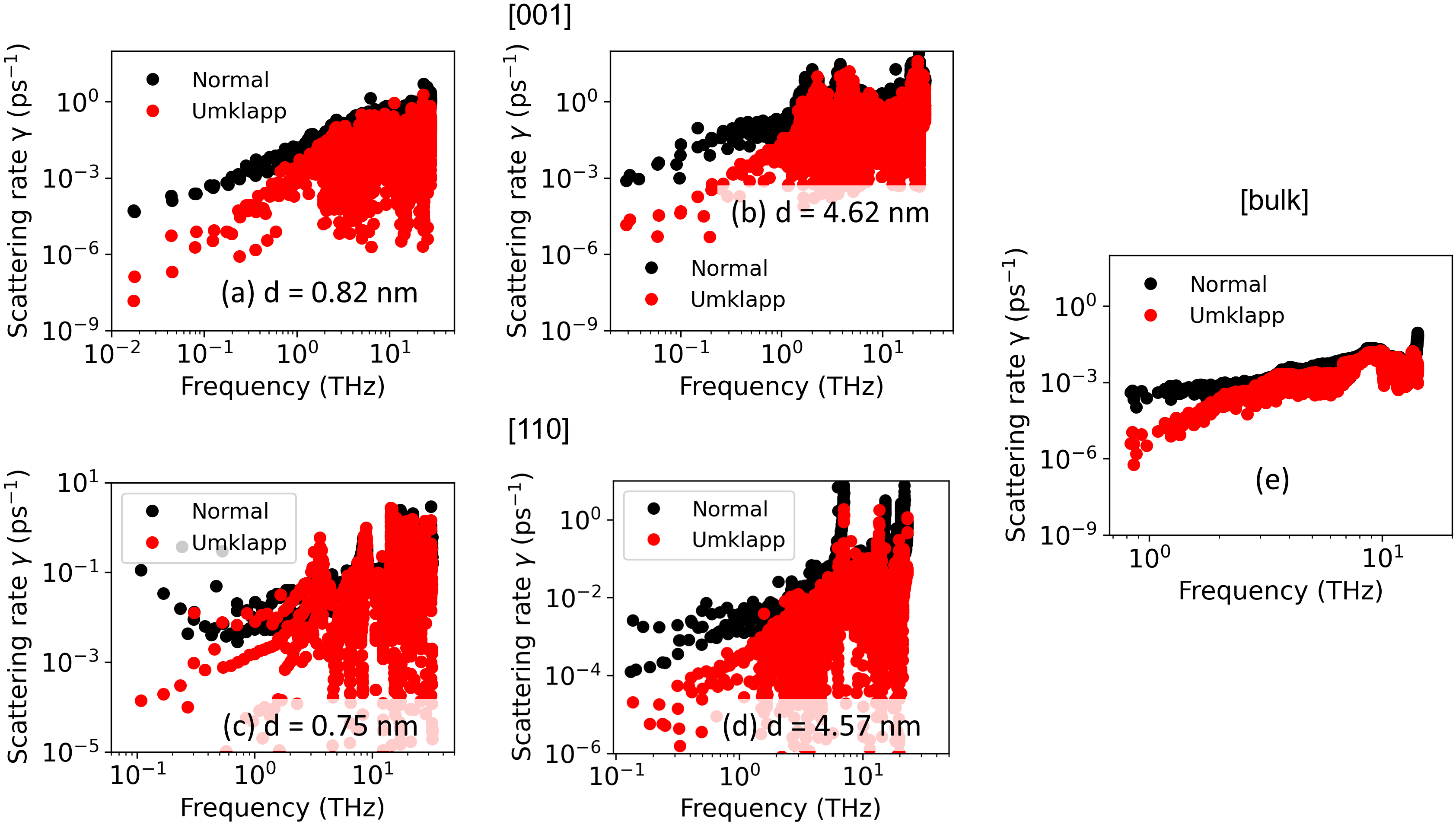}
    \caption{Calculated Normal (N) and Umklapp (U) phonon-scattering rates as functions of frequency for representative Si NWs and bulk Si. Panels (a) and (b) show [001]-oriented NWs with \(d=0.82\) nm and \(d=4.62\) nm, respectively; panels (c) and (d) show [110]-oriented NWs with \(d=0.75\) nm and \(d=4.57\) nm, respectively; panel (e) shows bulk Si. The smaller-diameter NWs are in the strongly confined regime below \(d_c\), whereas the larger-diameter NWs are close to their corresponding critical diameters.}
    \label{fig:scattering}
\end{figure}

Figure ~\ref{fig:scattering} compares the calculated Normal and Umklapp scattering rates for representative NWs in the strongly confined and near-critical diameter regimes. For the strongly confined [001]- (Fig.~\ref{fig:scattering} (a,b)) and [110]-oriented NWs (Fig.~\ref{fig:scattering} (c,d)), the Normal scattering rates are substantially larger than the corresponding Umklapp rates over the low-frequency range. As the diameter increases from the ultrathin regime toward the near-critical diameter, the difference between the Normal and Umklapp scattering rates decreases. This trend reflects the confinement-induced evolution of the phonon spectrum and the associated increase in the phase space available for Umklapp processes as the optical-like subbands shift to lower frequency. For diameters above \(d_c\), the relative influence of boundary and surface scattering is expected to decrease as the surface-to-volume ratio is reduced; nevertheless, these effects remain relevant throughout the sub-10 nm diameter range considered here. Fully anharmonic scattering-rate calculations above \(d_c\) are computationally prohibitive within the present framework. As a reference, the bulk-Si rates are shown in Fig.~\ref{fig:scattering}(e).

\begin{figure}[!tb]
    \centering
    \includegraphics[width=\linewidth]{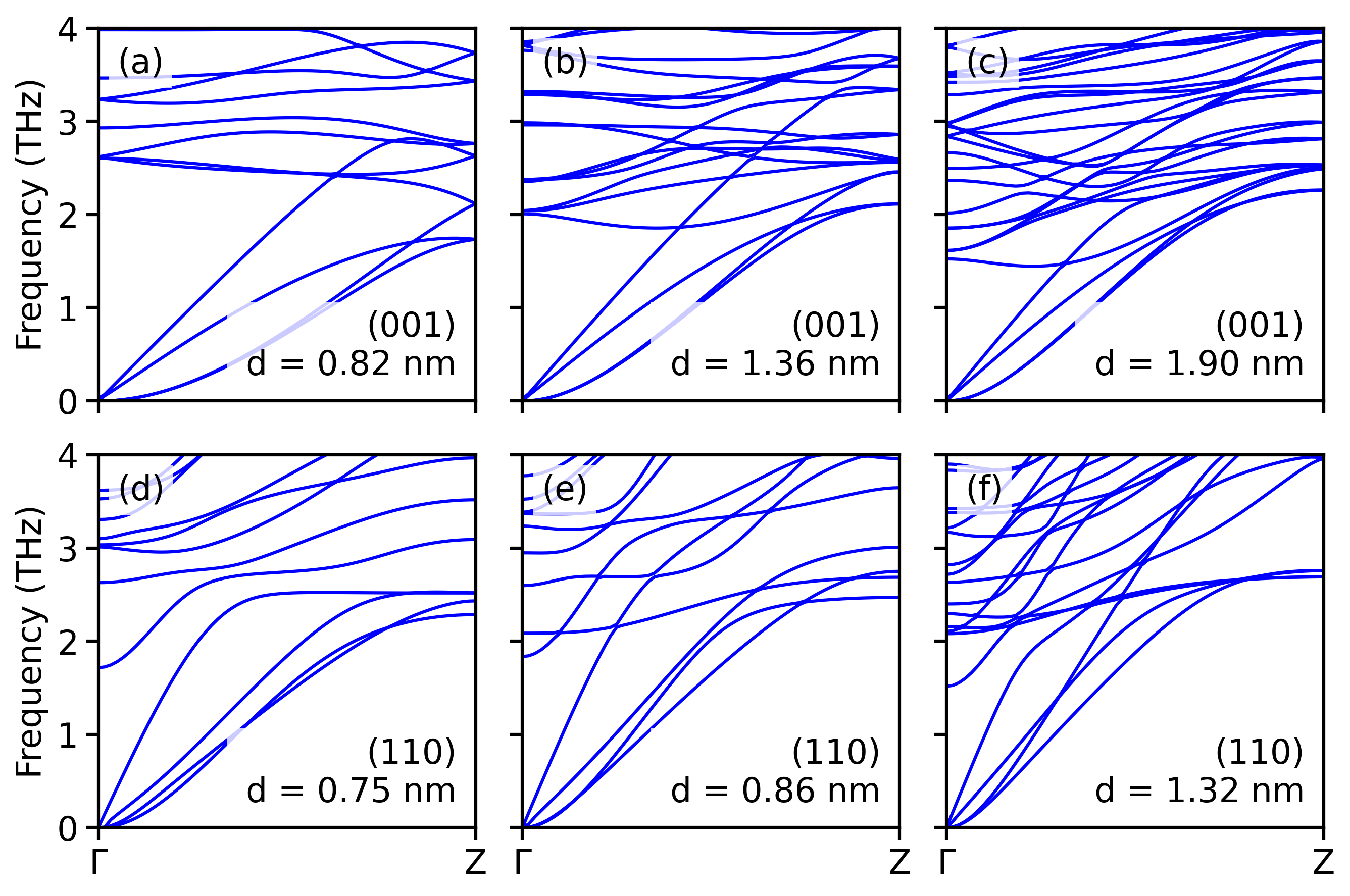}
    \caption{Phonon dispersion relations calculated for ultrathin Si NWs with different diameters. (a) [001]-oriented NW, diameter 0.82 nm; (b) [001], 1.36 nm; (c) [001], 1.90 nm; (d) [110]-oriented NW, diameter 0.75 nm; (e) [110], 0.86 nm; (f) [110], 1.32 nm.}
    \label{fig:bands}
\end{figure}

To gain more insights into the origin of the nonmonotonic diameter dependence of $\kappa$, we analyze phonon dispersions for both [001] and [110] Si NWs across a range of diameters, as shown in Fig.~\ref{fig:bands}. In agreement with Zhou \textit{et al.}~\cite{zhou2017nonmonotonic}, the surface-related low-lying optical branches systematically shift downward, e.g., by $\approx$ 1 THz when $d$ increases from 0.82 nm (Fig.~\ref{fig:bands} (a)) to 1.90 nm (Fig.~\ref{fig:bands} (c)). For low-frequency heat-carrying phonons ($\omega \lesssim 2$ THz), resistive Umklapp scattering occurs primarily via three-phonon absorption processes of the type acoustic + acoustic $\rightarrow$ surface optical (SO). The confinement-induced upward shift of the SO bands in the ultrathin NWs increases the gap between the acoustic and optical branches, largely reducing the Umklapp scattering rates ($\frac{1}{\tau_U}$). Nevertheless, at such low temperatures, Umklapp scattering among the acoustic branches is strongly suppressed, while normal scattering processes remain allowed and relatively higher ($\tau_N^{-1} \gg \tau_U^{-1}$) (see Fig.~\ref{fig:scattering}(a)). This dominance of momentum-conserving normal processes over resistive Umklapp scattering enables partial hydrodynamic phonon flow, thereby overcoming strong boundary scattering and yielding a thermal conductivity higher than expected for pure diffuse boundary-limited transport.

As the diameter increases from the ultrathin regime toward the near-critical range, the optical-like subbands shift downward in energy, increasing the phase space available for Umklapp processes. This trend is consistent with the reduced separation between the calculated Normal and Umklapp scattering rates in Fig.~\ref{fig:scattering}. Although boundary scattering weakens relative to the bulk as the diameter increases, it remains relevant over the sub-10 nm diameter range. The observed minimum in \(\kappa_{\mathrm{eff}}\) therefore reflects the combined effects of confinement-modified phonon structure, evolving anharmonic scattering phase space, and surface-limited propagation. For diameters above \(d_c\), \(\kappa_{\mathrm{eff}}\) increases with diameter in the present calculations. This increase is consistent with the reduced relative influence of surface and boundary scattering as the surface-to-volume ratio decreases; however, because fully anharmonic scattering rates were not systematically calculated beyond \(d_c\), we do not attribute the post-\(d_c\) trend to a specific Normal--Umklapp scattering crossover.

We next consider the cryogenic transport regime at \(T=10\) K. The effective thermal conductivity remains nonmonotonic with diameter for both crystallographic orientations. At this temperature, Bose–Einstein statistics strongly restrict the thermally active spectrum to low-frequency phonons, whereas high-frequency phonons have negligible occupation. The diameter dependence is therefore interpreted in terms of the confinement-modified low-frequency phonon spectrum, including the number and dispersion of thermally active subbands and their transmission through the finite device. Resistive Umklapp scattering is strongly suppressed for the thermally occupied modes and is not invoked as the origin of the decrease in \(\kappa_{\mathrm{eff}}\) toward \(d_c\). In the ultrathin limit, radial confinement discretizes the phonon spectrum into a small number of low-frequency acoustic-like subbands. The finite cross section and surface termination modify the transmission of these modes, and the finite channel length means that the calculated conductance should not be interpreted as a bulk diffusive limit. We therefore describe this regime as finite-length quantum phonon transport with few-mode or quasi-one-dimensional character, rather than as unqualified quasi-ballistic transport.

\begin{figure}[!tb]
    \centering
    \includegraphics[width=1\linewidth]{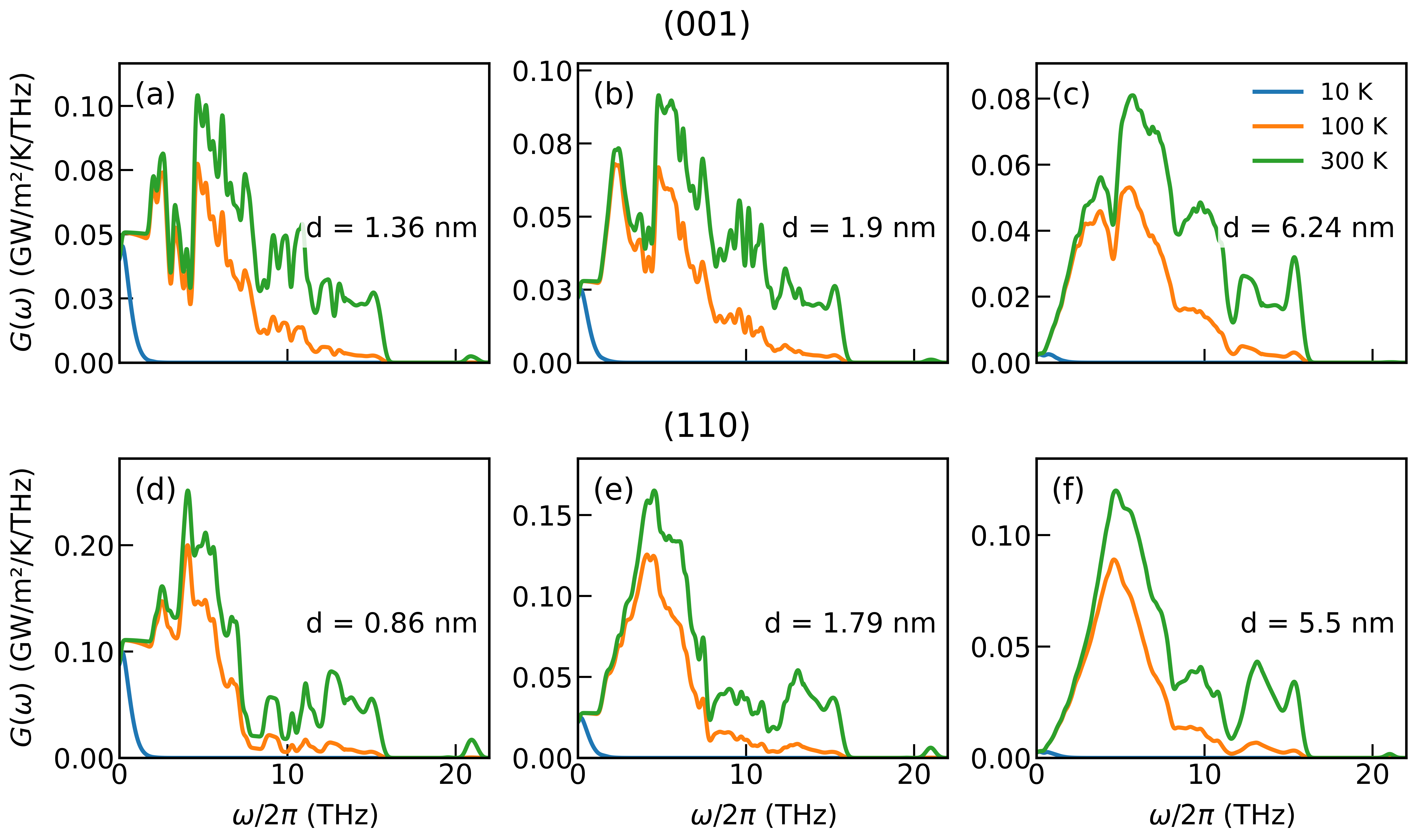}
    \caption{Spectral thermal conductance per unit cross-sectional area versus frequency for ultrathin silicon NWs at temperatures 10 K, 100 K, and 300 K (blue, orange, green curves). Top row: [001]-oriented NWs; bottom row: [110]-oriented NWs. Panels (a) and (b) show diameters below the critical value for [001], panels (d) and (e) for [110], while panels (c) and (f) correspond to the room-temperature critical diameters, \(d_c^{300\mathrm{K}}\approx6.24\) nm for [001] and \(\approx5.50\) nm for [110].}
    \label{fig:spectral}
\end{figure}

To separate the harmonic quantum-statistical contribution from the anharmonic correction, Fig. S7 compares harmonic Landauer and fully anharmonic NEGF spectral conductances for representative ultrathin [001]- and [110]-oriented Si NWs. At 10 K, both calculations show that heat transport is concentrated in the lowest-frequency confined acoustic-like modes. Table S1 summarizes the corresponding effective thermal conductivities. Over the 10–100 K range, anharmonic phonon–phonon scattering produces a finite but moderate suppression of \(\kappa_{\mathrm{eff}}\), while the harmonic calculation establishes the quantum-statistical baseline.

Figure~\ref{fig:spectral} shows the spectral thermal conductance of [001]-oriented (a--c) and [110]-oriented (d--f) Si NWs for three diameters at 10\,K, 100\,K, and 300\,K. Classical molecular dynamics simulations typically do not capture this behavior, as they follow Boltzmann statistics and excite all modes equally regardless of temperature or frequency.~\cite{liang2023mechanisms,wang2023quantum,xu2023accurate,gu2021thermal} For the smallest NW diameters, the spectral thermal conductance shows pronounced contributions from low-frequency acoustic-like branches [Fig. 5(a,d)]. As the diameter increases toward \(d_c\), additional low-frequency confined subbands emerge [Fig. 5(b,e)]. At cryogenic temperature, their contributions are governed by Bose–Einstein occupation, confinement-modified dispersion, and finite-device transmission under surface-limited propagation. Resistive Umklapp scattering is suppressed for the thermally occupied modes; the harmonic–anharmonic comparison in Fig. S6 instead demonstrates a finite anharmonic correction to \(\kappa_{\mathrm{eff}}\). As temperature rises, higher-frequency phonons become thermally populated and increasingly contribute to the conductance.

To further examine the cryogenic transport regime, Fig. S5 presents temperature-dependent spectral thermal conductances for representative ultrathin [001]- and [110]-oriented NWs. At 5–10 K, the heat-carrying spectrum is restricted primarily to the lowest-frequency confined acoustic-like modes, whereas higher-frequency phonons are strongly suppressed following Bose–Einstein occupation. As temperature increases, the thermally active frequency window broadens, and additional confined subbands progressively contribute to the conductance. This behavior indicates a few-mode, quasi-one-dimensional transport character in the ultrathin limit.

\section{Conclusion}
In summary, we used a DFT-trained machine-learning potential with anharmonic quantum NEGF calculations to investigate thermal transport in ultrathin silicon NWs. The calculated \(\kappa_{\mathrm{eff}}(d)\) exhibits nonmonotonic behavior: \(\kappa_{\mathrm{eff}}\) first decreases with increasing \(d\), reaches temperature-dependent minima at critical diameters \(d_c^{300\,\mathrm{K}}\) and \(d_c^{10\,\mathrm{K}}\), and then increases as \(d\) increases further. At high temperatures, this behavior results from competition between Normal and Umklapp scattering processes. At cryogenic temperatures, Bose--Einstein statistics restrict heat transport to low-frequency, confined acoustic-like modes. Comparison with the harmonic Landauer limit shows that anharmonic phonon--phonon scattering produces a finite correction to the effective thermal conductivity. In ultrathin NWs, radial confinement discretizes the phonon spectrum into a few low-lying acoustic-like subbands, giving the cryogenic regime a few-mode, quasi-one-dimensional character. Because all simulated devices have a channel length of 10 nm, the reported conductivities are finite-length effective values rather than bulk-limit NW thermal conductivities. The MLP--NEGF framework developed here provides fully anharmonic, quantum-mechanical, and atomically resolved predictions of thermal transport in silicon NWs. It offers atomic-scale insight into confinement-driven nonmonotonic behavior and can support the rational design of nanoelectronic and thermoelectric devices for thermal-management applications.

\begin{acknowledgments}
This work was supported by the Defense Advanced Research Projects Agency (DARPA) under the Thermal Modeling of Nanoscale Transistors (Thermonat) program. Computations were performed using the Partnership for an Advanced Computing Environment (PACE) cluster at Georgia Institute of Technology and the Anvil cluster at Purdue University through allocation MCH240072.

\end{acknowledgments}

\bibliography{references.bib}

@article{casimir1938note,
  title={Note on the conduction of heat in crystals},
  author={Casimir, HBG},
  journal={Physica},
  volume={5},
  number={6},
  pages={495--500},
  year={1938},
  doi={10.1016/S0031-8914(38)80162-2},
  publisher={Elsevier}
}

@article{xu2025critical,
  title={{Critical Size Transitions in Silicon Nanowires: Amorphization, Phonon Hydrodynamics, and Thermal Conductivity}},
  author={Xu, Ke and Li, Yuan and Ding, Dongliang and Liang, Ting and Wu, Jianyang and Xu, Jianbin},
  journal={The Journal of Physical Chemistry Letters},
  volume={16},
  number={33},
  pages={8580--8587},
  year={2025},
  doi={10.1021/acs.jpclett.5c01802},
  publisher={ACS Publications}
}

@article{zhou2017nonmonotonic,
  title={{Nonmonotonic diameter dependence of thermal conductivity of extremely thin Si nanowires: competition between hydrodynamic phonon flow and boundary scattering}},
  author={Zhou, Yanguang and Zhang, Xiaoliang and Hu, Ming},
  journal={Nano Letters},
  volume={17},
  number={2},
  pages={1269--1276},
  year={2017},
  doi={10.1021/acs.nanolett.6b05113},
  publisher={ACS Publications}
}

@article{gireesan2020diameter,
  title={{Diameter-dependent thermal conductivity of ultrathin GaP nanowires: a molecular dynamics study}},
  author={Gireesan, Subash and Torres, Pol and Alvarez, F Xavier and Bobbert, Peter A},
  journal={Physical Review B},
  volume={101},
  number={2},
  pages={024307},
  year={2020},
  doi={10.1103/PhysRevB.101.024307},
  publisher={APS}
}

@article{cahill2003nanoscale,
  title={Nanoscale thermal transport},
  author={Cahill, David G and Ford, Wayne K and Goodson, Kenneth E and Mahan, Gerald D and Majumdar, Arun and Maris, Humphrey J and Merlin, Roberto and Phillpot, Simon R},
  journal={Journal of Applied Physics},
  volume={93},
  number={2},
  pages={793--818},
  year={2003},
  doi={10.1063/1.1524305},
  publisher={American Institute of Physics}
}

@article{nomura2022review,
  title={Review of thermal transport in phononic crystals},
  author={Nomura, Masahiro and Anufriev, Roman and Zhang, Zhongwei and Maire, Jeremie and Guo, Yangyu and Yanagisawa, Ryoto and Volz, Sebastian},
  journal={Materials Today Physics},
  volume={22},
  pages={100613},
  year={2022},
  doi={10.1016/j.mtphys.2022.100613},
  publisher={Elsevier}
}

@article{zhang2018theoretical,
  title={A theoretical review on interfacial thermal transport at the nanoscale},
  author={Zhang, Ping and Yuan, Peng and Jiang, Xiong and Zhai, Siping and Zeng, Jianhua and Xian, Yaoqi and Qin, Hongbo and Yang, Daoguo},
  journal={Small},
  volume={14},
  number={2},
  pages={1702769},
  year={2018},
  doi={10.1002/smll.201702769},
  publisher={Wiley Online Library}
}

@article{beardo2025nanoscale,
  title={Nanoscale confinement of phonon flow and heat transport},
  author={Beardo, Albert and Chen, Weinan and McBennett, Brendan and Karimzadeh Sabet, Tara and Nelson, Emma E and Culman, Theodore H and Kapteyn, Henry C and Knobloch, Joshua L and Murnane, Margaret M and Dabo, Ismaila},
  journal={npj Computational Materials},
  volume={11},
  number={1},
  pages={172},
  year={2025},
  doi={10.1038/s41524-025-01593-7},
  publisher={Nature Publishing Group UK London}
}

@article{hochbaum2010semiconductor,
  title={Semiconductor nanowires for energy conversion},
  author={Hochbaum, Allon I and Yang, Peidong},
  journal={Chemical Reviews},
  volume={110},
  number={1},
  pages={527--546},
  year={2010},
  doi={10.1021/cr900075v},
  publisher={ACS Publications}
}

@article{swinkels2018nanowires,
  title={Nanowires for heat conversion},
  author={Swinkels, Milo Yaro and Zardo, Ilaria},
  journal={Journal of Physics D: Applied Physics},
  volume={51},
  number={35},
  pages={353001},
  year={2018},
  doi={10.1088/1361-6463/aad25f},
  publisher={IOP Publishing}
}

@article{dasgupta201425th,
  title={25th anniversary article: semiconductor nanowires--synthesis, characterization, and applications},
  author={Dasgupta, Neil P and Sun, Jianwei and Liu, Chong and Brittman, Sarah and Andrews, Sean C and Lim, Jongwoo and Gao, Hanwei and Yan, Ruoxue and Yang, Peidong},
  journal={Advanced Materials},
  volume={26},
  number={14},
  pages={2137--2184},
  year={2014},
  doi={10.1002/adma.201305929},
  publisher={Wiley Online Library}
}

@article{hochbaum2008enhanced,
  title={Enhanced thermoelectric performance of rough silicon nanowires},
  author={Hochbaum, Allon I and Chen, Renkun and Delgado, Raul Diaz and Liang, Wenjie and Garnett, Erik C and Najarian, Mark and Majumdar, Arun and Yang, Peidong},
  journal={Nature},
  volume={451},
  number={7175},
  pages={163--167},
  year={2008},
  doi={10.1038/nature06381},
  publisher={Nature Publishing Group UK London}
}

@article{chen2019thermoelectrics,
  title={Thermoelectrics of nanowires},
  author={Chen, Renkun and Lee, Jaeho and Lee, Woochul and Li, Deyu},
  journal={Chemical Reviews},
  volume={119},
  number={15},
  pages={9260--9302},
  year={2019},
  doi={10.1021/acs.chemrev.8b00627},
  publisher={ACS Publications}
}

@article{shakouri2006nanoscale,
  title={Nanoscale thermal transport and microrefrigerators on a chip},
  author={Shakouri, Ali},
  journal={Proceedings of the IEEE},
  volume={94},
  number={8},
  pages={1613--1638},
  year={2006},
  doi={10.1109/JPROC.2006.879787},
  publisher={IEEE}
}

@article{chang2023nanowire,
  title={Nanowire-based integrated photonics for quantum information and quantum sensing},
  author={Chang, Jin and Gao, Jun and Esmaeil Zadeh, Iman and Elshaari, Ali W and Zwiller, Val},
  journal={Nanophotonics},
  volume={12},
  number={3},
  pages={339--358},
  year={2023},
  doi={10.1515/nanoph-2022-0652},
  publisher={De Gruyter}
}

@article{smidstrup2019quantumatk,
  title={QuantumATK: an integrated platform of electronic and atomic-scale modelling tools},
  author={Smidstrup, S{\o}ren and Markussen, Troels and Vancraeyveld, Pieter and Wellendorff, Jess and Schneider, Julian and Gunst, Tue and Verstichel, Brecht and Stradi, Daniele and Khomyakov, Petr A and Vej-Hansen, Ulrik G and others},
  journal={Journal of Physics: Condensed Matter},
  volume={32},
  number={1},
  pages={015901},
  year={2019},
  doi={10.1088/1361-648X/ab4007},
  publisher={IOP Publishing}
}

@article{hahn2023thermal,
  title={{Thermal transport in H-terminated ultrathin [110] Si nanowires: A first principles study}},
  author={Hahn, Konstanze R and Melis, Claudio and Bernardini, Fabio and Paulatto, Lorenzo and Colombo, Luciano},
  journal={The European Physical Journal Plus},
  volume={138},
  number={11},
  pages={1028},
  year={2023},
  doi={10.1140/epjp/s13360-023-04659-x},
  publisher={Springer}
}

@article{boukai2008silicon,
  title={Silicon nanowires as efficient thermoelectric materials},
  author={Boukai, Akram I and Bunimovich, Yuri and Tahir-Kheli, Jamil and Yu, Jen-Kan and Goddard Iii, William A and Heath, James R},
  journal={Nature},
  volume={451},
  number={7175},
  pages={168--171},
  year={2008},
  doi={10.1038/nature06458},
  publisher={Nature Publishing Group UK London}
}

@article{ponomareva2007thermal,
  title={Thermal conductivity in thin silicon nanowires: phonon confinement effect},
  author={Ponomareva, Inna and Srivastava, Deepak and Menon, Madhu},
  journal={Nano Letters},
  volume={7},
  number={5},
  pages={1155--1159},
  year={2007},
  doi={10.1021/nl062823d},
  publisher={ACS Publications}
}

@article{donadio2010temperature,
  title={Temperature dependence of the thermal conductivity of thin silicon nanowires},
  author={Donadio, Davide and Galli, Giulia},
  journal={Nano Letters},
  volume={10},
  number={3},
  pages={847--851},
  year={2010},
  doi={10.1021/nl903268y},
  publisher={ACS Publications}
}

@article{guo2020quantum,
  title={Quantum mechanical modeling of anharmonic phonon-phonon scattering in nanostructures},
  author={Guo, Yangyu and Bescond, Marc and Zhang, Zhongwei and Luisier, Mathieu and Nomura, Masahiro and Volz, Sebastian},
  journal={Physical Review B},
  volume={102},
  number={19},
  pages={195412},
  year={2020},
  doi={10.1103/PhysRevB.102.195412},
  publisher={APS}
}

@article{guo2021anharmonic,
  title={{Anharmonic phonon-phonon scattering at the interface between two solids by nonequilibrium Green's function formalism}},
  author={Guo, Yangyu and Zhang, Zhongwei and Bescond, Marc and Xiong, Shiyun and Nomura, Masahiro and Volz, Sebastian},
  journal={Physical Review B},
  volume={103},
  number={17},
  pages={174306},
  year={2021},
  doi={10.1103/PhysRevB.103.174306},
  publisher={APS}
}

@article{luisier2012atomistic,
  title={Atomistic modeling of anharmonic phonon-phonon scattering in nanowires},
  author={Luisier, Mathieu},
  journal={Physical Review B},
  volume={86},
  number={24},
  pages={245407},
  year={2012},
  doi={10.1103/PhysRevB.86.245407},
  publisher={APS}
}

@article{dong2024molecular,
  title={{Molecular dynamics simulations of heat transport using machine-learned potentials: A mini-review and tutorial on GPUMD with neuroevolution potentials}},
  author={Dong, Haikuan and Shi, Yongbo and Ying, Penghua and Xu, Ke and Liang, Ting and Wang, Yanzhou and Zeng, Zezhu and Wu, Xin and Zhou, Wenjiang and Xiong, Shiyun and Chen, Shunda and Fan, Zheyong},
  journal={Journal of Applied Physics},
  volume={135},
  number={16},
  year={2024},
  url={https://doi.org/10.1063/5.0200833},
  publisher={AIP Publishing}
}

@article{wang2007quantum,
  title={Quantum thermal transport from classical molecular dynamics},
  author={Wang, Jian-Sheng},
  journal={Physical Review Letters},
  volume={99},
  number={16},
  pages={160601},
  year={2007},
  doi={10.1103/PhysRevLett.99.160601},
  publisher={APS}
}

@book{datta2005quantum,
  title={Quantum transport: atom to transistor},
  author={Datta, Supriyo},
  year={2005},
  publisher={Cambridge University Press}
}

@article{zeng2018nanoscale,
  title={{Nanoscale thermal transport: Theoretical method and application}},
  author={Zeng, Yu-Jia and Liu, Yue-Yang and Zhou, Wu-Xing and Chen, Ke-Qiu},
  journal={Chinese Physics B},
  volume={27},
  number={3},
  pages={036304},
  year={2018},
  doi={10.1088/1674-1056/27/3/036304},
  publisher={IOP Publishing}
}

@article{chen2018thermal,
  title={{Thermal engineering in low-dimensional quantum devices: A tutorial review of nonequilibrium Green's function methods}},
  author={Chen, Xiaobin and Liu, Yizhou and Duan, Wenhui},
  journal={Small Methods},
  volume={2},
  number={6},
  pages={1700343},
  year={2018},
  doi={10.1002/smtd.201700343},
  publisher={Wiley Online Library}
}

@article{fan2021neuroevolution,
  title={{Neuroevolution machine learning potentials: Combining high accuracy and low cost in atomistic simulations and application to heat transport}},
  author={Fan, Zheyong and Zeng, Zezhu and Zhang, Cunzhi and Wang, Yanzhou and Song, Keke and Dong, Haikuan and Chen, Yue and Ala-Nissila, Tapio},
  journal={Physical Review B},
  volume={104},
  number={10},
  pages={104309},
  year={2021},
  doi={10.1103/PhysRevB.104.104309},
  publisher={APS}
}

@article{unruh2022gaussian,
  title={{Gaussian approximation potential for amorphous Si: H}},
  author={Unruh, Davis and Meidanshahi, Reza Vatan and Goodnick, Stephen M and Cs{\'a}nyi, G{\'a}bor and Zim{\'a}nyi, Gergely T},
  journal={Physical Review Materials},
  volume={6},
  number={6},
  pages={065603},
  year={2022},
  doi={10.1103/PhysRevMaterials.6.065603},
  publisher={APS}
}

@article{fan2022gpumd,
  title={{GPUMD: A package for constructing accurate machine-learned potentials and performing highly efficient atomistic simulations}},
  author={Fan, Zheyong and Wang, Yanzhou and Ying, Penghua and Song, Keke and Wang, Junjie and Wang, Yong and Zeng, Zezhu and Xu, Ke and Lindgren, Eric and Rahm, J Magnus and others},
  journal={The Journal of Chemical Physics},
  volume={157},
  number={11},
  year={2022},
  url={https://doi.org/10.1063/5.0106617},
  publisher={AIP Publishing}
}

@article{he2012microscopic,
  title={Microscopic origin of the reduced thermal conductivity of silicon nanowires},
  author={He, Yuping and Galli, Giulia},
  journal={Physical Review Letters},
  volume={108},
  number={21},
  pages={215901},
  year={2012},
  doi={10.1103/PhysRevLett.108.215901},
  publisher={APS}
}

@article{li2003thermal,
  title={Thermal conductivity of individual silicon nanowires},
  author={Li, Deyu and Wu, Yiying and Kim, Philip and Shi, Li and Yang, Peidong and Majumdar, Arun},
  journal={Applied Physics Letters},
  volume={83},
  number={14},
  pages={2934--2936},
  year={2003},
  doi={10.1063/1.1616981},
  publisher={American Institute of Physics}
}

@article{lee2016thermal,
  title={{Thermal transport in silicon nanowires at high temperature up to 700 K}},
  author={Lee, Jaeho and Lee, Woochul and Lim, Jongwoo and Yu, Yi and Kong, Qiao and Urban, Jeffrey J and Yang, Peidong},
  journal={Nano Letters},
  volume={16},
  number={7},
  pages={4133--4140},
  year={2016},
  doi={10.1021/acs.nanolett.6b00956},
  publisher={ACS Publications}
}

@article{dames2004theoretical,
  title={{Theoretical phonon thermal conductivity of Si/Ge superlattice nanowires}},
  author={Dames, Chris and Chen, Gang},
  journal={Journal of Applied Physics},
  volume={95},
  number={2},
  pages={682--693},
  year={2004},
  doi={10.1063/1.1631734},
  publisher={American Institute of Physics}
}

@article{rashid2018effect,
  title={Effect of confinement on anharmonic phonon scattering and thermal conductivity in pristine silicon nanowires},
  author={Rashid, Zahid and Zhu, Liyan and Li, Wu},
  journal={Physical Review B},
  volume={97},
  number={7},
  pages={075441},
  year={2018},
  doi={10.1103/PhysRevB.97.075441},
  publisher={APS}
}

@article{yan2016phonon,
  title={Phonon transport at the interfaces of vertically stacked graphene and hexagonal boron nitride heterostructures},
  author={Yan, Zhequan and Chen, Liang and Yoon, Mina and Kumar, Satish},
  journal={Nanoscale},
  volume={8},
  number={7},
  pages={4037--4046},
  year={2016},
  doi={10.1039/c5nr06818e},
  publisher={Royal Society of Chemistry}
}

@article{kresse1996efficiency,
  title={Efficiency of ab-initio total energy calculations for metals and semiconductors using a plane-wave basis set},
  author={Kresse, Georg and Furthm{\"u}ller, J{\"u}rgen},
  journal={Computational Materials Science},
  volume={6},
  number={1},
  pages={15--50},
  year={1996},
  doi={10.1016/0927-0256(96)00008-0},
  publisher={Elsevier}
}

@article{kresse1996efficient,
  title={Efficient iterative schemes for ab initio total-energy calculations using a plane-wave basis set},
  author={Kresse, Georg and Furthm{\"u}ller, J{\"u}rgen},
  journal={Physical Review B},
  volume={54},
  number={16},
  pages={11169},
  year={1996},
  doi={10.1103/PhysRevB.54.11169},
  publisher={APS}
}

@article{perdew1996generalized,
  title={Generalized gradient approximation made simple},
  author={Perdew, John P and Burke, Kieron and Ernzerhof, Matthias},
  journal={Physical Review Letters},
  volume={77},
  number={18},
  pages={3865},
  year={1996},
  doi={10.1103/PhysRevLett.77.3865},
  publisher={APS}
}

@article{blochl1994projector,
  title={Projector augmented-wave method},
  author={Bl{\"o}chl, Peter E},
  journal={Physical Review B},
  volume={50},
  number={24},
  pages={17953},
  year={1994},
  doi={10.1103/PhysRevB.50.17953},
  publisher={APS}
}

@article{monkhorst1976special,
  title={{Special points for Brillouin-zone integrations}},
  author={Monkhorst, Hendrik J and Pack, James D},
  journal={Physical Review B},
  volume={13},
  number={12},
  pages={5188},
  year={1976},
  doi={10.1103/PhysRevB.13.5188},
  publisher={APS}
}

@article{wang2006nonequilibrium,
  title={{Nonequilibrium Green’s function approach to mesoscopic thermal transport}},
  author={Wang, Jian-Sheng and Wang, Jian and Zeng, Nan},
  journal={Physical Review B},
  volume={74},
  number={3},
  pages={033408},
  year={2006},
  doi={10.1103/PhysRevB.74.033408},
  publisher={APS}
}

@article{mingo2006anharmonic,
  title={Anharmonic phonon flow through molecular-sized junctions},
  author={Mingo, N},
  journal={Physical Review B},
  volume={74},
  number={12},
  pages={125402},
  year={2006},
  doi={10.1103/PhysRevB.74.125402},
  publisher={APS}
}

@article{yan2016role,
  title={{The role of interfacial electronic properties on phonon transport in two-dimensional MoS$_2$ on metal substrates}},
  author={Yan, Zhequan and Chen, Liang and Yoon, Mina and Kumar, Satish},
  journal={ACS Applied Materials \& Interfaces},
  volume={8},
  number={48},
  pages={33299--33306},
  year={2016},
  doi={10.1021/acsami.6b10608},
  publisher={ACS Publications}
}

@article{liang2022abnormally,
  title={Abnormally high thermal conductivity in fivefold twinned diamond nanowires},
  author={Liang, Ting and Xu, Ke and Han, Meng and Yao, Yimin and Zhang, Zhisen and Zeng, Xiaoliang and Xu, Jianbin and Wu, Jianyang},
  journal={Materials Today Physics},
  volume={25},
  pages={100705},
  year={2022},
  doi={10.1016/j.mtphys.2022.100705},
  publisher={Elsevier}
}

@article{liang2023mechanisms,
  title={Mechanisms of temperature-dependent thermal transport in amorphous silica from machine-learning molecular dynamics},
  author={Liang, Ting and Ying, Penghua and Xu, Ke and Ye, Zhenqiang and Ling, Chao and Fan, Zheyong and Xu, Jianbin},
  journal={Physical Review B},
  volume={108},
  number={18},
  pages={184203},
  year={2023},
  doi={10.1103/PhysRevB.108.184203},
  publisher={APS}
}

@article{fan2019homogeneous,
  title={Homogeneous nonequilibrium molecular dynamics method for heat transport and spectral decomposition with many-body potentials},
  author={Fan, Zheyong and Dong, Haikuan and Harju, Ari and Ala-Nissila, Tapio},
  journal={Physical Review B},
  volume={99},
  number={6},
  pages={064308},
  year={2019},
  doi={10.1103/PhysRevB.99.064308},
  publisher={APS}
}

@article{wang2023quantum,
  title={Quantum-corrected thickness-dependent thermal conductivity in amorphous silicon predicted by machine learning molecular dynamics simulations},
  author={Wang, Yanzhou and Fan, Zheyong and Qian, Ping and Caro, Miguel A and Ala-Nissila, Tapio},
  journal={Physical Review B},
  volume={107},
  number={5},
  pages={054303},
  year={2023},
  doi={10.1103/PhysRevB.107.054303},
  publisher={APS}
}

@article{xu2023accurate,
  title={Accurate prediction of heat conductivity of water by a neuroevolution potential},
  author={Xu, Ke and Hao, Yongchao and Liang, Ting and Ying, Penghua and Xu, Jianbin and Wu, Jianyang and Fan, Zheyong},
  journal={The Journal of Chemical Physics},
  volume={158},
  number={20},
  pages={204114},
  year={2023},
  doi={10.1063/5.0147039},
  publisher={AIP Publishing}
}

@article{gu2021thermal,
  title={{Thermal conductivity prediction by atomistic simulation methods: Recent advances and detailed comparison}},
  author={Gu, Xiaokun and Fan, Zheyong and Bao, Hua},
  journal={Journal of Applied Physics},
  volume={130},
  number={21},
  pages={210902},
  year={2021},
  doi={10.1063/5.0069175},
  publisher={AIP Publishing}
}

@article{zaccone2025phonon,
  title={Phonon-confinement theory of thermal conductivity in ultrathin silicon films},
  author={Zaccone, Alessio},
  journal={Journal of Applied Physics},
  volume={138},
  number={22},
  pages={225305},
  year={2025},
  doi={10.1063/5.0304896},
  publisher={AIP Publishing}
}

@article{machida2020phonon,
  title={Phonon hydrodynamics and ultrahigh--room-temperature thermal conductivity in thin graphite},
  author={Machida, Yo and Matsumoto, Nayuta and Isono, Takayuki and Behnia, Kamran},
  journal={Science},
  volume={367},
  number={6475},
  pages={309--312},
  year={2020},
  doi={10.1126/science.aaz8043},
  publisher={American Association for the Advancement of Science}
}

@article{liu2022non,
  title={Non-monotonic thickness dependent and anisotropic in-plane thermal transport in layered titanium trisulphide},
  author={Liu, C and Lu, P and Li, D and Zhao, Y and Hao, M},
  journal={Materials Today Nano},
  volume={17},
  pages={100165},
  year={2022},
  doi={10.1016/j.mtnano.2021.100165},
  publisher={Elsevier}
}

@article{yuan2018nonmonotonic,
  title={{Nonmonotonic thickness-dependence of in-plane thermal conductivity of few-layered MoS$_2$: 2.4 to 37.8 nm}},
  author={Yuan, Pengyu and Wang, Ridong and Wang, Tianyu and Wang, Xinwei and Xie, Yangsu},
  journal={Physical Chemistry Chemical Physics},
  volume={20},
  number={40},
  pages={25752--25761},
  year={2018},
  doi={10.1039/C8CP02858C},
  publisher={Royal Society of Chemistry}
}

@article{wu2020anomalous,
  title={{Anomalous layer thickness dependent thermal conductivity of Td-WTe$_2$ through first-principles calculation}},
  author={Wu, Chao and Liu, Chenhan and Tao, Yi and Zhang, Yan and Chen, Yunfei},
  journal={Physics Letters A},
  volume={384},
  number={30},
  pages={126751},
  year={2020},
  doi={10.1016/j.physleta.2020.126751},
  publisher={Elsevier}
}

@article{siemens2010quasi,
  title={{Quasi-ballistic thermal transport from nanoscale interfaces observed using ultrafast coherent soft X-ray beams}},
  author={Siemens, Mark E and Li, Qing and Yang, Ronggui and Nelson, Keith A and Anderson, Erik H and Murnane, Margaret M and Kapteyn, Henry C},
  journal={Nature materials},
  volume={9},
  number={1},
  pages={26--30},
  year={2010},
  doi={10.1038/nmat2568},
  publisher={Nature Publishing Group UK London}
}

@article{fan2022improving,
  title={Improving the accuracy of the neuroevolution machine learning potential for multi-component systems},
  author={Fan, Zheyong},
  journal={Journal of Physics: Condensed Matter},
  volume={34},
  number={12},
  pages={125902},
  year={2022},
  doi={10.1088/1361-648X/ac462b},
  publisher={IOP Publishing}
}

@article{gao2022catalyst,
  title={Catalyst-free synthesis of sub-5 nm silicon nanowire arrays with massive lattice contraction and wide bandgap},
  author={Gao, Sen and Hong, Sanghyun and Park, Soohyung and Jung, Hyun Young and Liang, Wentao and Lee, Yonghee and Ahn, Chi Won and Byun, Ji Young and Seo, Juyeon and Hahm, Myung Gwan and Kim, Hyehee and Kim, Kiwoong and Yi, Yeonjin and Wang, Hailong and Upmanyu, Moneesh and Lee, Sung-Goo and Homma, Yoshikazu and Terrones, Humberto and Jung, Yung Joon},
  journal={Nature Communications},
  volume={13},
  number={1},
  pages={3467},
  year={2022},
  doi={10.1038/s41467-022-31174-x},
  publisher={Nature Publishing Group UK London}
}

@article{singh2026revealing,
  title={{Revealing Phonon Bridge Effect for Amorphous vs Crystalline Metal--Silicide Layers at Si/Ti Interfaces by a Machine Learning Potential}},
  author={Singh, Mayur and Patra, Lokanath and Zhang, Chengyang and MacDougall, Gregory J and Datta, Suman and Cahill, David G and Kumar, Satish},
  journal={ACS Applied Electronic Materials},
  volume={8},
  number={6},
  pages={2373--2383},
  year={2026},
  doi={10.1021/acsaelm.5c02592},
  publisher={ACS Publications}
}

@article{yue2025nanoparticle,
  title={Nanoparticle-assisted phonon transport modulation in Si/Ge heterostructures using neuroevolution potential machine learning models},
  author={Yue, Jincheng and Chen, Rongkun and Ma, Dengke and Hu, Shiqian},
  journal={Chinese Physics Letters},
  volume={42},
  number={3},
  pages={036301},
  year={2025},
  publisher={Chinese Physical Society and IOP Publishing Ltd}
}

@article{yin2026accelerated,
  title={Accelerated phonon transport calculations for nanostructures: Combining neuroevolution potentials and compressed sensing},
  author={Yin, Fei and Liu, Shixian and Dong, Yiming and Barinov, AA and Xu, Ke and Khvesyuk, VI},
  journal={Journal of Applied Physics},
  volume={139},
  number={13},
  year={2026},
  publisher={AIP Publishing}
}

@article{wu2026atomic,
  title={Atomic-scale bonding configurations and phonon-bridge-mediated interfacial thermal transport in GaN/SiC},
  author={Wu, Ning and Wu, Zhe and Yang, Rui and Huang, Dian and Tang, Guihua and Liu, Zhigang},
  journal={Physical Review B},
  volume={113},
  number={19},
  pages={195308},
  year={2026},
  publisher={APS}
}

@article{cheng2025revealing,
  title={Revealing phonon signature of dislocations in silicon carbide using machine-learning interatomic potential},
  author={Cheng, Mo and Jiang, Xuanyu and Zhang, Haoming and Pi, Xiaodong and Yang, Deren and Deng, Tianqi},
  journal={Applied Physics Letters},
  volume={127},
  number={24},
  year={2025},
  publisher={AIP Publishing}
}

\end{document}



\title{Supplementary Information: \\ Anharmonic Quantum Transport Analysis of Thermal Transport Anomalies in Ultrathin Silicon Nanowires} 

\author{Lokanath Patra}
\affiliation{%
George W. Woodruff School of Mechanical Engineering, Georgia Institute of Technology, Atlanta, GA 30332, USA}%

\author{Mayur Pratap Singh}
\affiliation{%
George W. Woodruff School of Mechanical Engineering, Georgia Institute of Technology, Atlanta, GA 30332, USA}%

\author{Satish Kumar}
\email{satish.kumar@me.gatech.edu} 
\affiliation{%
George W. Woodruff School of Mechanical Engineering, Georgia Institute of Technology, Atlanta, GA 30332, USA}%

\maketitle


\section{Computational Methods}

\subsection{Density Functional Theory Simulations}
Density functional theory structural relaxations were performed using the Vienna Ab initio Simulation Package (VASP)~\cite{kresse1996efficiency,kresse1996efficient} with the projector augmented-wave (PAW) method~\cite{blochl1994projector}. The exchange-correlation energy was described within the generalized-gradient approximation using the Perdew-Burke-Ernzerhof (PBE) functional~\cite{perdew1996generalized}. Monkhorst-Pack~\cite{monkhorst1976special} $\textbf{k}$-point grids of $12 \times 12 \times 12$ and $1 \times 1 \times 10$ were used to sample the Brillouin zones of bulk Si and nanowires (NWs), respectively. The electronic total energy and ionic forces were converged to $1 \times 10^{-6}$ eV and 0.01 eV/\AA, respectively. To model the NWs, a vacuum region of approximately 15 \AA~was introduced along the $x$ and $y$ directions to eliminate periodic interactions.

\subsection{Neuroevolution Potential Training}
A neuroevolution potential (NEP)~\cite{fan2021neuroevolution} was used to perform the NEGF and MD simulations for this work. We used an initial NEP radial and angular cutoff of 5 angstroms to train the NEP, while leaving all other weighting parameters at their default values. The training data consist of 390 samples of crystalline positions, forces, energies, and virial stress for crystalline Si, amorphous Si, compounds, and SiH interfaces, generated previously using a Gaussian Approximation Potential.~\cite{unruh2022gaussian} After convergence at 300000 generations, the NEP had error RMSEs of 6.4 meV/atom, 157 meV/Å, and 31 meV/atom for energies, forces, and virials, respectively. We produced a test dataset of small SiH nanowires with various diameters and found force errors below 112 meV/Å, which increased confidence in the potential.

\begin{figure}[!htb]
    \centering
    \includegraphics[width=1\linewidth]{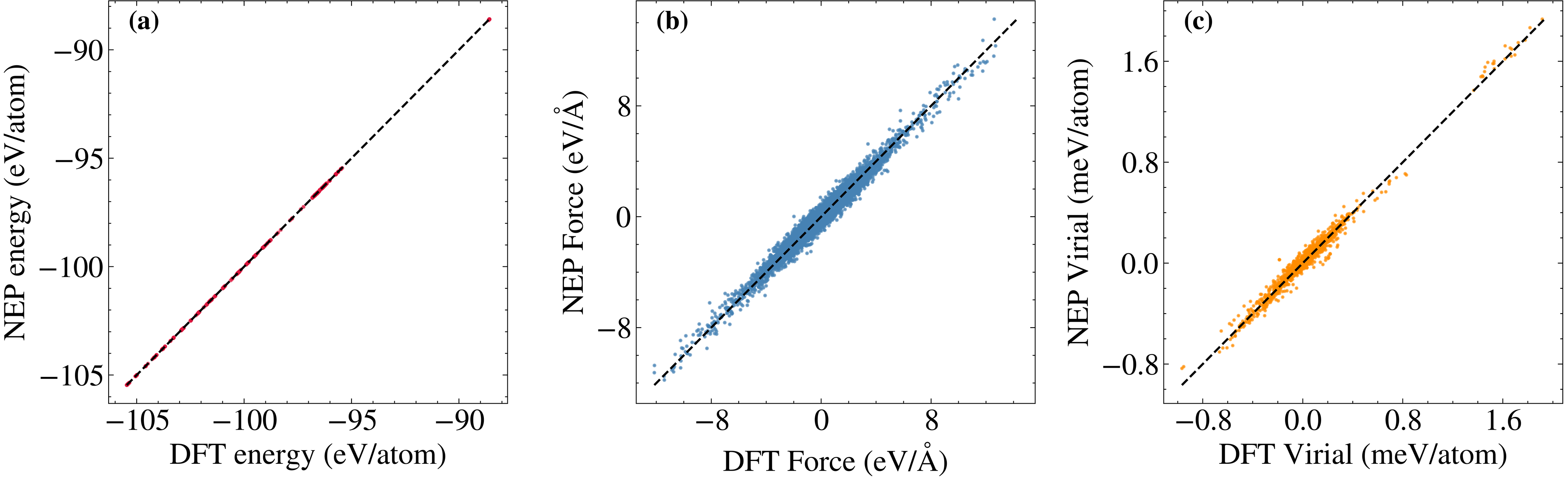}
    \caption{Training accuracy of the neuroevolution potential (NEP): a) energy root-mean-square error (RMSE), b) force RMSE, and c) virial RMSE, clustered by system type in the training dataset.}
    \label{fig:RDF}
\end{figure}

The training database includes crystalline Si, amorphous Si, Si-containing compounds, and Si--H interface configurations, with DFT reference energies, atomic forces, and virial stresses used in the fit. The potential reproduces the overall bulk-Si phonon dispersion, including the low-frequency acoustic branches, while deviations remain for portions of the optical branches. Such residual optical-branch deviations are common for machine-learning interatomic potentials fitted to finite configuration databases and may be reduced by including additional distorted and off-equilibrium configurations in future training sets [\cite{xu2025critical,yue2025nanoparticle,yin2026accelerated}]. Since the low-temperature conductance considered here is dominated by low-frequency acoustic-like modes, the optical-branch deviations do not alter the qualitative cryogenic interpretation. At higher temperatures, where optical modes can contribute more substantially, the results are interpreted in terms of the directly calculated transport trends rather than individual optical-branch features.

To further assess the structural and vibrational transferability of the NEP, we compared the bulk-Si phonon density of states calculated using DFT and NEP force constants, as well as radial distribution functions obtained from DFT-MD and NEP-MD simulations for bulk Si and a representative H-passivated [001]-oriented Si nanowire. As shown in Fig.~\ref{fig:RDF}(a), the NEP captures the principal features of the DFT phonon density of states, particularly in the low- and intermediate-frequency regions. Some differences remain in the high-frequency optical region, consistent with the deviations observed in the optical branches of the phonon dispersion. Similar residual discrepancies in high-frequency optical modes have been reported for machine-learning interatomic potentials developed for Si and related systems.~\cite{xu2025critical,yin2026accelerated,wu2026atomic,cheng2025revealing}

Figures \ref{fig:RDF}(b) and \ref{fig:RDF}(c) compare the normalized Si--Si radial distribution functions for bulk Si and the [001]-oriented Si nanowire, respectively. In both systems, the NEP-MD and DFT-MD results show closely matching positions and overall shapes for the principal coordination-shell peaks. Figure \ref{fig:RDF}(d) compares the normalized Si--H radial distribution function for the H-passivated nanowire. The NEP reproduces the first-neighbor Si--H peak, indicating that it captures the local Si--H bonding environment, although differences remain in the longer-range correlation features. Together with the bulk phonon-dispersion comparison, these results provide structural and harmonic vibrational validation of the NEP for bulk Si and H-passivated Si nanowires.

\begin{figure}
    \centering
    \includegraphics[width=\linewidth]{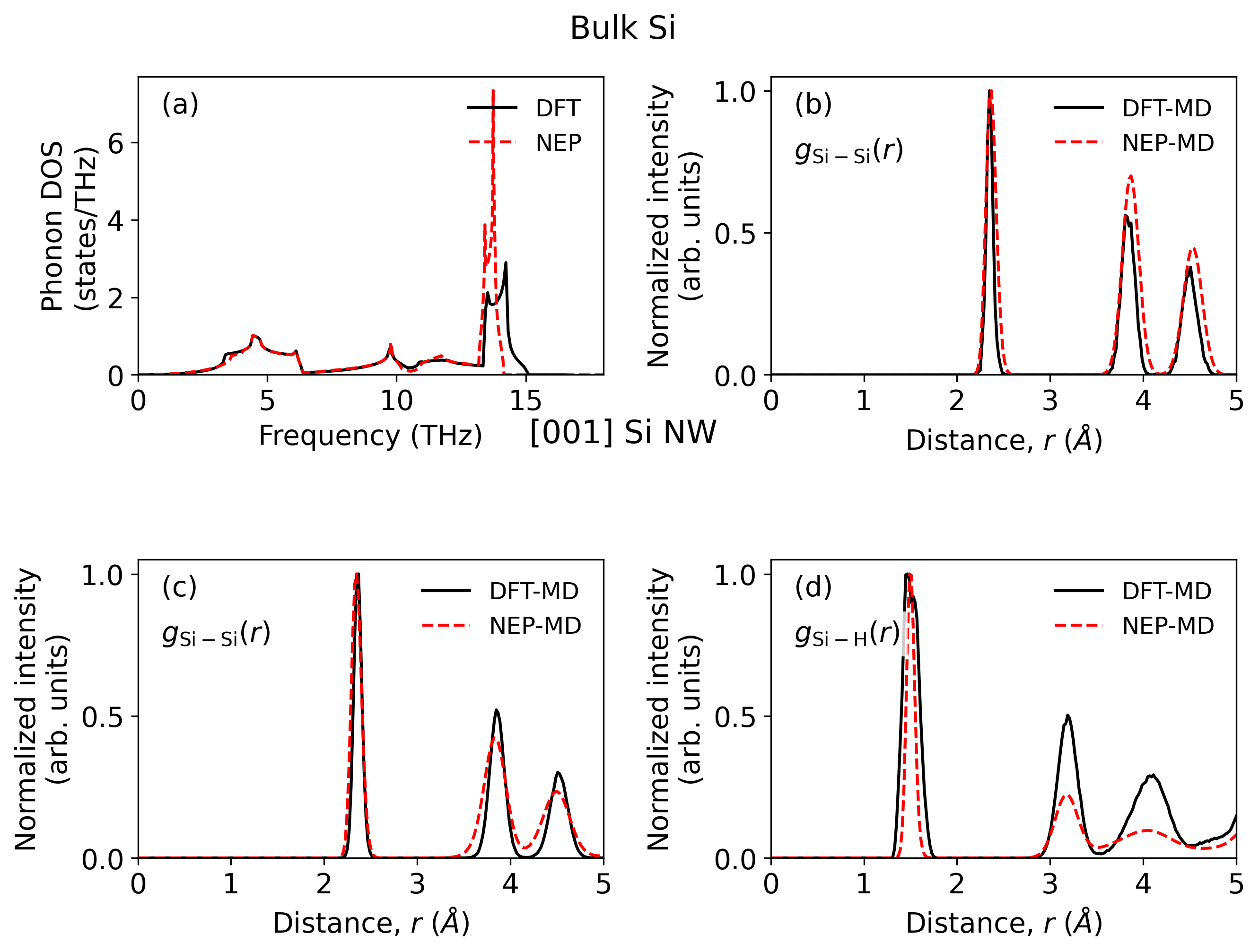}
    \caption{Comparison of DFT and NEP results for bulk Si and an H-passivated [001] Si nanowire. (a) Bulk-Si phonon density of states. Normalized radial distribution functions from DFT-MD and NEP-MD for (b) bulk Si--Si, (c) nanowire Si--Si, and (d) nanowire Si--H pairs.}
    \label{fig:RDF}
\end{figure}

To further assess the transport performance of the trained NEP, we calculated the lattice thermal conductivity of bulk Si by solving the phonon Boltzmann transport equation (BTE). Harmonic and third-order anharmonic force constants were generated using the NEP, which was interfaced with the \texttt{phono3py} package through the \texttt{calorine} Python package. The calculations employed a \(4\times4\times4\) bulk-Si supercell and a \(30\times30\times30\) \(q\)-point mesh for the BTE evaluation. 

\begin{figure}[!t]
    \centering
    \includegraphics[width=0.5\linewidth]{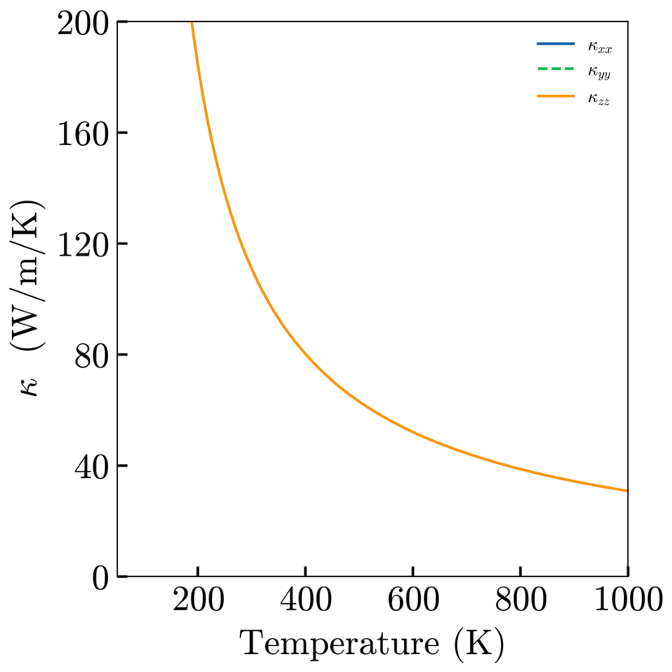}
    \caption{Temperature dependence of the lattice thermal conductivity of bulk Si calculated by solving the phonon Boltzmann transport equation using NEP-derived harmonic and third-order anharmonic force constants.}
    \label{fig:BTE}
\end{figure}

Figure~\ref{fig:BTE} shows the temperature dependence of the calculated bulk-Si lattice thermal conductivity. The three Cartesian components, \(\kappa_{xx}\), \(\kappa_{yy}\), and \(\kappa_{zz}\), are nearly identical, as expected for cubic silicon. At 300 K, the calculated conductivity is \(110~\mathrm{W\,m^{-1}\,K^{-1}}\), compared with the commonly reported experimental value of approximately \(150~\mathrm{W\,m^{-1}\,K^{-1}}\). The moderate underestimation at room temperature is consistent with previous work showing that force errors in machine-learned potentials can lead to underestimated lattice thermal conductivities through additional effective phonon scattering. This benchmark therefore confirms that the NEP captures the expected qualitative bulk-Si transport behavior and provides a suitable computationally efficient force-field description for the present comparative nanowire transport calculations.

\subsection{Nonequilibrium Green's Function Theory}
We extended our previously developed atomistic Green’s function (AGF) code for ballistic heat transport~\cite{yan2016phonon,yan2016role} by explicitly including anharmonic phonon–phonon scattering, following the approach of Guo \textit{et al.}~\cite{guo2020quantum,guo2021anharmonic}.

In matrix notation, the retarded Green's function of the device is given by~\cite{guo2020quantum,guo2021anharmonic,wang2006nonequilibrium,mingo2006anharmonic,luisier2012atomistic}
\begin{equation}
    \mathbf{G}^{r}(\omega; \mathbf{q}_\perp) = \bigl[\omega^{2}\mathbf{I} - \boldsymbol{\Phi}( \mathbf{q}_\perp) - \boldsymbol{\Sigma}^{r}_{L}(\omega; \mathbf{q}_\perp) - \boldsymbol{\Sigma}^{r}_{R}(\omega; \mathbf{q}_\perp) - \boldsymbol{\Sigma}^{r}_{S}(\omega; \mathbf{q}_\perp)\bigr]^{-1},
\end{equation}
where $(\omega; \mathbf{q}_\perp)$ are the frequency and wave-vector along the transport and transverse directions, respectively; $\mathbf{I}$ and $\boldsymbol{\Phi}$ are the unit matrix and second-order force constant matrix, respectively; $\boldsymbol{\Sigma}^{r}_{L}(\omega)$ and $\boldsymbol{\Sigma}^{r}_{R}(\omega)$ represent the retarded self-energy matrix for left and right contacts, respectively. The anharmonic phonon-phonon scattering in the device region is represented by $\boldsymbol{\Sigma}^{r}_{S}(\omega)$. The greater/lesser Green's function needs to be calculated for anharmonic heat transport as:
\begin{equation}
    \mathbf{G}^{>,<}(\omega; \mathbf{q}_\perp) = \mathbf{G}^r(\omega; \mathbf{q}_\perp) \, \Sigma^{>,<}(\omega; \mathbf{q}_\perp) \, \mathbf{G}^a(\omega; \mathbf{q}_\perp)
\end{equation}
where $\mathbf{G}^a(\omega)$ is the advanced Green's function of the device and is the Hermitian conjugate of the retarded Green's function,
\begin{equation}
\mathbf{G}^a(\omega; \mathbf{q}_\perp) = \bigl[\mathbf{G}^r(\omega; \mathbf{q}_\perp)\bigr]^\dagger
\end{equation}
The greater/lesser self-energies also have contributions from both the contacts and anharmonic interaction, i.e.
\begin{equation}
    \boldsymbol{\Sigma}^{>,<}(\omega) = \boldsymbol{\Sigma}^{>,<}_L(\omega) + \boldsymbol{\Sigma}^{>,<}_R(\omega) +\boldsymbol{\Sigma}^{>,<}_S(\omega)
\end{equation}

The retarded scattering self-energy can be calculated from the greater/lesser self-energy as
\begin{equation}
    \boldsymbol{\Sigma}^r(\omega; \mathbf{q}_\perp) = \frac{\boldsymbol{\Sigma}^>(\omega; \mathbf{q}_\perp) - \boldsymbol{\Sigma}^<(\omega; \mathbf{q}_\perp)}{2} 
                               + \mathrm{i} \, \mathcal{P} \int_{-\infty}^{\infty} \frac{\mathrm{d}\omega'}{2\pi} 
                                 \frac{\boldsymbol{\Sigma}^>(\omega'; \mathbf{q}_\perp) - \boldsymbol{\Sigma}^<(\omega'); \mathbf{q}_\perp}{\omega - \omega'},
\end{equation}
where the second term on the right-hand side is the Cauchy principal value integral, which is commonly neglected to reduce computational complexity.

Next, the greater/lesser self-energy matrix is calculated as
\begin{equation}
\begin{aligned}
\boldsymbol{\Sigma}^{>,<}_{ij,S}(ll',\omega;\mathbf{q}_\perp) &= 
\frac{i\hbar}{2} \sum_{l_1\dots l_4} \sum_{j_1\dots j_4}
\int_{-\infty}^{\infty} \frac{\mathrm{d}\omega'}{2\pi}\;
\boldsymbol{\Phi}_{i j_1 j_2}(l l_1 l_2;\mathbf{q}_\perp',\mathbf{q}_\perp-\mathbf{q}_\perp')\;
\boldsymbol{\Phi}_{j j_3 j_4}(l' l_3 l_4;\mathbf{q}_\perp',\mathbf{q}_\perp-\mathbf{q}_\perp') \\
&\quad \times \;
\boldsymbol{G}^{>,<}_{j_1 j_4}(l_1 l_4;\omega',\mathbf{q}_\perp')\;
\boldsymbol{G}^{>,<}_{j_2 j_3}(l_2 l_3;\omega-\omega',\mathbf{q}_\perp-\mathbf{q}_\perp').
\end{aligned}
\end{equation}
where $\boldsymbol{\Phi}_{l l_1 l_2}^{i j_1 j_2}$ are the third-order force constants matrices, the subscripts $l(l_1, l_2, l_3, l_4)$ indicate atomic indices, and the superscripts $i,j$ indicate Cartesian coordinates ($x,y,z$)

In the presence of anharmonic scattering, the transmission function $T(\omega; \mathbf{q}_\perp)$ is computed using the Caroli formula, which accounts for both elastic and inelastic phonon processes through the self-consistent Green's functions:
\begin{equation}
    T(\omega; \mathbf{q}_\perp) = \mathrm{Tr} \left[ \boldsymbol{\Gamma}_L(\omega; \mathbf{q}_\perp) \, \mathbf{G}^r(\omega; \mathbf{q}_\perp) \, \boldsymbol{\Gamma}_R(\omega; \mathbf{q}_\perp) \, \mathbf{G}^a(\omega; \mathbf{q}_\perp) \right],
\end{equation}
where $\boldsymbol{\Gamma}_{L/R}(\omega; \mathbf{q}_\perp) = i \left[ \boldsymbol{\Sigma}^{r}_{L/R}(\omega; \mathbf{q}_\perp) - \boldsymbol{\Sigma}^{a}_{L/R}(\omega; \mathbf{q}_\perp) \right]/2$ are the broadening matrices from the left and right contacts, and the trace is taken over the device degrees of freedom. This expression incorporates the effects of anharmonic scattering via the scattering self-energy $\boldsymbol{\Sigma}^{r}_{S}$ embedded in $\mathbf{G}^{r/a}$. For systems with transverse periodicity, the total transmission is obtained by summing (or integrating) over the transverse Brillouin zone and, as appropriate, normalizing by the number of transverse unit cells.

In the linear response regime, the net heat current $J$ flowing through a nanoscale junction or interface under a small temperature difference $\Delta T = T_L - T_R$ (with $T_L$ and $T_R$ the temperatures of the left and right reservoirs, respectively) is given by the Landauer--B\"{u}ttiker-like expression for phonons:

\begin{equation}
J = \frac{1}{2\pi} \int_0^\infty \hbar \omega \, T(\omega) \, \left[ n_B(\omega, T_L) - n_B(\omega, T_R) \right] \, d\omega,
\label{eq:heat_current}
\end{equation}
where $T(\omega)$ is the energy- and mode-resolved transmission probability (transmission function), $n_B(\omega, T) = \left[ \exp(\hbar \omega / k_B T) - 1 \right]^{-1}$ is the Bose--Einstein distribution function, and the integral is performed over angular frequency $\omega$. In the limit of infinitesimal temperature bias ($\Delta T \to 0$, $T_L \approx T_R \approx T$), the heat current becomes proportional to $\Delta T$, and the thermal conductance $\kappa = J / \Delta T$ is obtained by differentiating with respect to temperature:

\begin{equation}
\kappa = \frac{\partial J}{\partial (\Delta T)} \bigg|_{\Delta T \to 0} = \frac{1}{2\pi} \int_0^\infty \hbar \omega \, T(\omega) \, \frac{\partial n_B(\omega, T)}{\partial T} \, d\omega.
\label{eq:total_conductance}
\end{equation}

The frequency-resolved (spectral) thermal conductance is then defined as the integrand per unit frequency interval:

\begin{equation}
G(\omega) = \frac{\hbar \omega}{2\pi} \, T(\omega) \, \frac{\partial n_B(\omega, T)}{\partial T}.
\label{eq:spectral_conductance}
\end{equation}

Using the standard derivative of the Bose function,

\begin{equation}
\frac{\partial n_B(\omega, T)}{\partial T} = \frac{1}{T} \, \frac{x \, e^x}{(e^x - 1)^2}, \qquad x = \frac{\hbar \omega}{k_B T},
\label{eq:bose_derivative}
\end{equation}

which yields the commonly used form

\begin{equation}
G(\omega) = \frac{\hbar \omega \, T(\omega)}{2\pi} \, \frac{x \, e^x}{(e^x - 1)^2} \, \frac{1}{T}.
\label{eq:g_omega_final}
\end{equation}

After normalization by the effective cross-sectional area $A$ of the system (yielding units of GW\,m$^{-2}$\,K$^{-1}$ per unit frequency), $G(\omega)$ represents the contribution of phonons at frequency $\omega$ to the total interfacial thermal conductance. To produce smooth, physically interpretable spectra suitable for comparison with experiments or classical simulations, Gaussian broadening with frequency-dependent widths is typically applied to $G(\omega)$ before plotting or integration, accounting for finite-lifetime effects, numerical discretization noise, and mode overlap in dense spectral regions.

\begin{figure}[!ht]
    \centering
    \includegraphics[width=\linewidth]{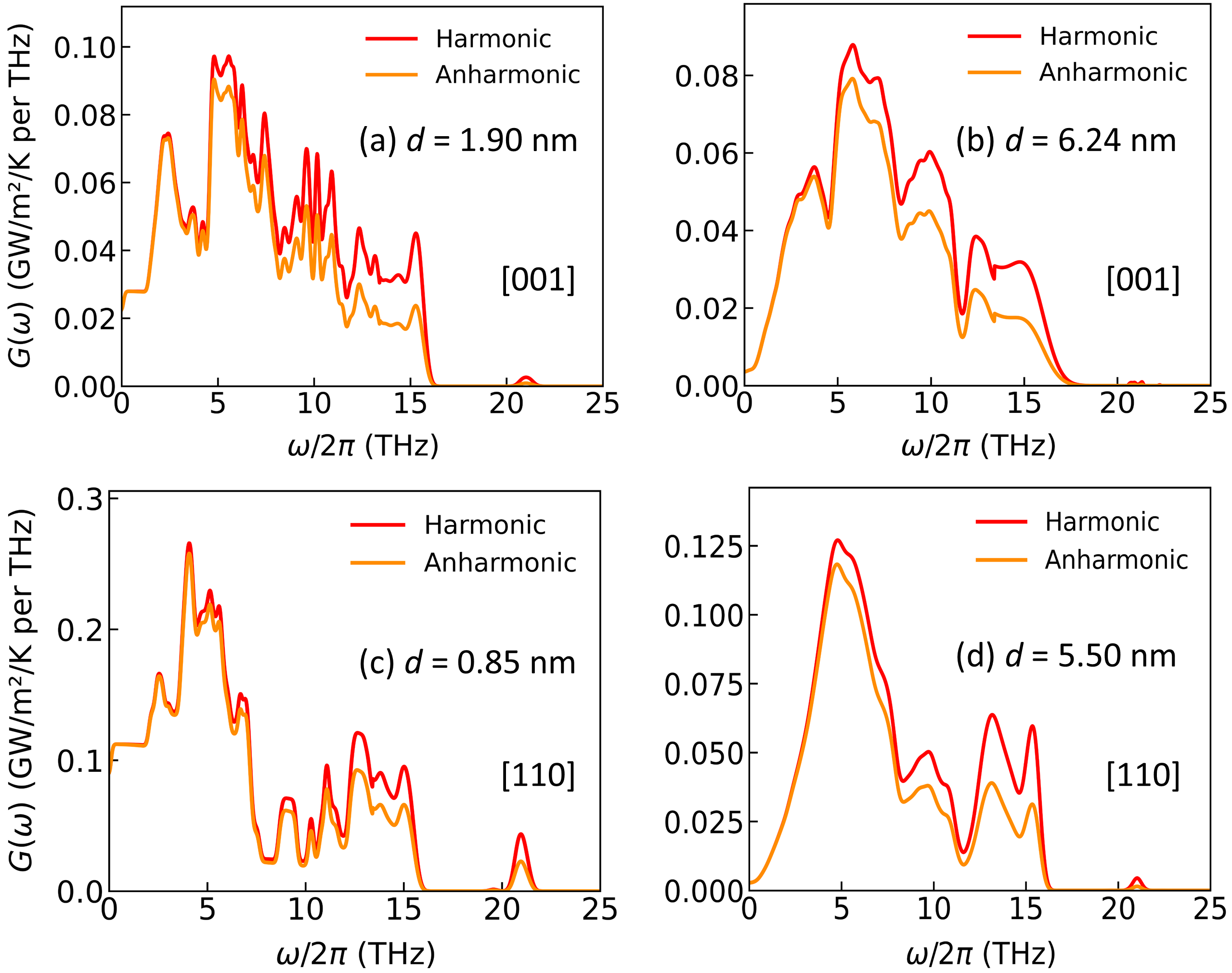}
    \caption{Spectral thermal conductance $G(\omega)$ obtained from harmonic and anharmonic NEGF simulations of silicon NWs at 300 K. Top panels (a) and (b) show [001]-oriented NWs with diameters (a) $d = 1.90$\,nm ($d < d_c$, thin) and (b) $d = 6.24$\,nm ($d = d_c$). Bottom panels (c) and (d) show [110]-oriented NWs with diameters (c) $d = 0.85$\,nm ($d < d_c$, ultrathin) and (d) $d = 5.50$\,nm ($d = d_c$). Inclusion of third-order anharmonic phonon scattering leads to progressively stronger suppression of $G(\omega)$, especially at higher frequencies and in thicker NWs.}
    \label{fig:placeholder}
\end{figure}

\subsection{Nonequilibrium Molecular Dynamics Simulations}
MD simulations were performed using GPUMD,~\cite{fan2022gpumd} and we calculated the thermal conductivity using homogeneous non-equilibrium MD (HNEMD) simulations.~\cite{fan2019homogeneous} In the HNEMD approach, an external driving force $F_i^{ext}$ is added to each atom $i$ considered in the simulation; this force has the form: 

\begin{equation}
    F_i^{ext}=E_iF_e+\sum_{j\neq i}{(\frac{\partial U_j}{\partial r_{ji}}}\ \bigotimes\ r_{ij})\bullet F_e
\end{equation}
where $E_i$ is the total energy of atom $i$. $U_i$ is the potential energy of atom $i$. $r_{ij}$ is the pairwise distance obtained from the subtraction of the positions $r_i$ and $r_j$ of atom $i$ and $j$, respectivelyThe summed tensor product represents the atomic virial stress of atom $i$ with respect to all other relevant atoms.. $F_e$ is an inverse length parameter that controls the magnitude of the driving force. The driving force will induce a non-equilibrium heat current $\langle J \rangle_{ne}$ which we can relate to $F_e$ via: 
\begin{equation}
    \frac{\left\langle J^\mu\left(t\right)\right\rangle_{\mathrm{ne}}}{TV}=\sum_{\nu}\kappa^{\mu\nu}F_\mathrm{e}^\nu
\end{equation}
where $\kappa^{\mu\nu}$ is the thermal conductivity tensor, T is the system temperature, and V is the system volume. The system's external temperature also needs to be controlled using a thermostat, such as a Nose-Hoover NVT. 

For HNEMD simulations, choosing $F_e$ is nontrivial because it must be sufficiently small to keep the system within the linear-response regime. Otherwise, the thermal conductivity will not converge during the simulation. After some testing, $F_e$ was set to 0.01 $\mu$Å$^{-1}$. The simulation protocol is as follows: we set the time step to 0.5 fs, then equilibrate the nanowire at room temperature using the Nose-Hoover NVT thermostat for 1 ns. We then maintain the temperature with the same NVT parameters, apply a driving force, and calculate the thermal conductivity using HNEMD over a 20 ns production run. We performed five independent simulations to obtain the room-temperature conductivity of 10-nm nanowires. We also tested for length effects at 30 nm and observed a 7\% modulation in measured average thermal conductivity. 

\section{Harmonic vs. Anharmonic NEGF Results}
To understand the effect of anharmonic phonon-phonon scattering, we have compared the spectral thermal conductance $G(\omega)$ of both [001] and [110] Si NWs, computed via harmonic and anharmonic NEGF simulations. The results reveal a diameter-dependent suppression when third-order phonon-phonon scattering is included. For both [001]-oriented NWs ($d = 1.90$\,nm below $d_c$ and $d = 6.24$\,nm at $d_c$) and [110]-oriented NWs ($d = 0.85$\,nm below $d_c$ and $d = 5.50$\,nm at $d_c$), anharmonic effects reduce $G(\omega)$ most strongly at higher frequencies. This frequency dependence stems from the rapid increase in three-phonon scattering rates with frequency, typically scaling as $\tau^{-1} \propto \omega^2$ or faster, because the phase space for energy- and momentum-conserving processes grows rapidly at high $\omega$. In thicker NWs approaching $d_c$, confinement flattens the phonon dispersion relations, lowering group velocities and increasing the phonon density of states. These changes enlarge the available phase space for anharmonic decay of high-frequency modes, resulting in significantly shorter mean free paths for those modes. In contrast, low-frequency acoustic phonons scatter much less and remain largely ballistic. These findings highlight the importance of third-order anharmonicity in the strong reduction of high-frequency thermal transport in silicon NWs.

\section{Length Dependence of the Effective Thermal Conductivity}

\begin{figure}[!htb]
    \centering
    \includegraphics[width=0.5\linewidth]{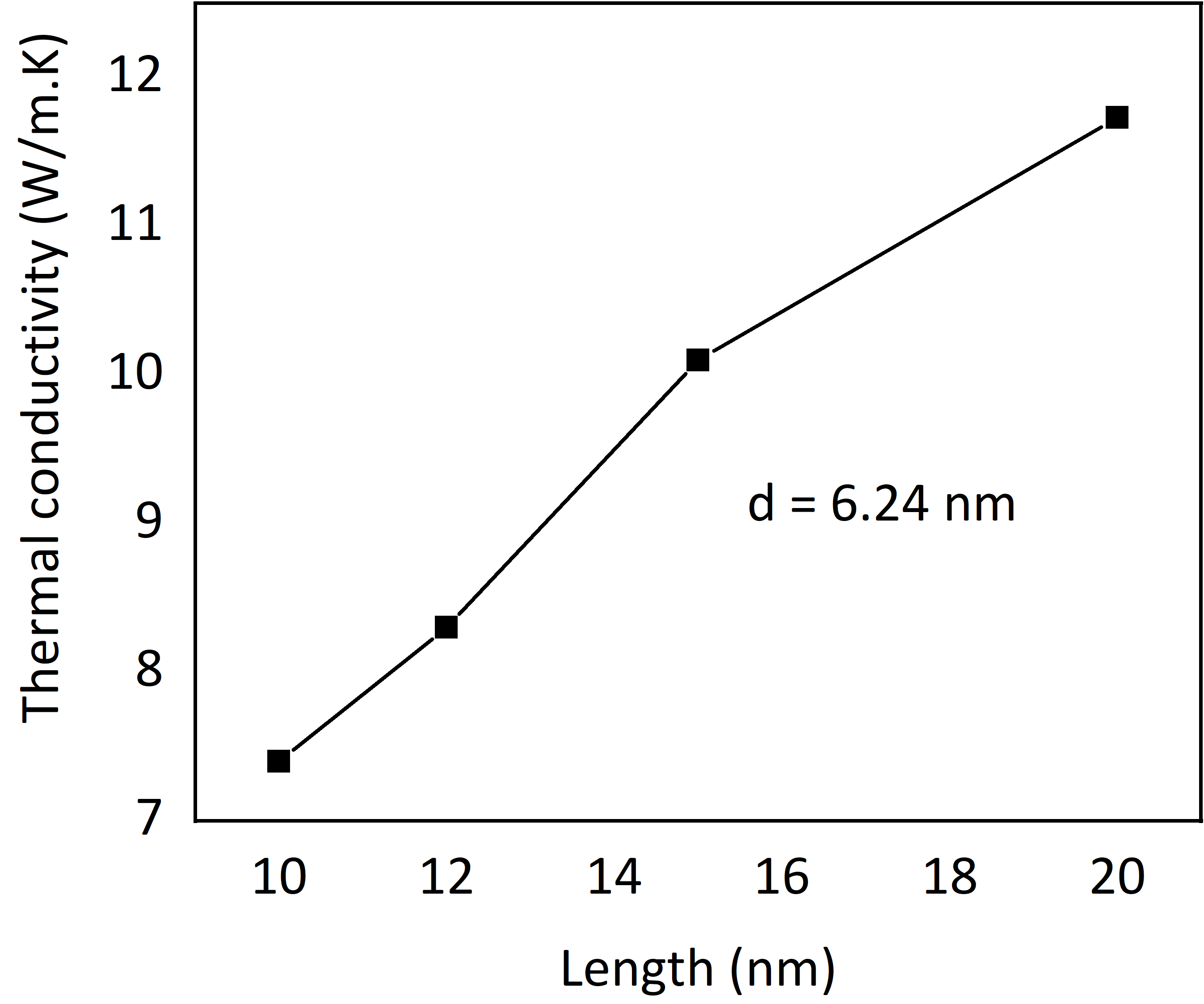}
    \caption{Length dependence of the effective thermal conductivity, \(\kappa_{\mathrm{eff}}=GL/A\), for the H-passivated [001]-oriented Si nanowire with \(d=6.24\) nm at 300 K. Fully anharmonic NEGF calculations were performed for channel lengths \(L=10\), 12, 15, and 20 nm. The increase in \(\kappa_{\mathrm{eff}}\) with \(L\) demonstrates that the calculated transport remains length-dependent over the investigated range. Thus, the \(L=10\) nm results in the main text are finite-length effective thermal conductivities rather than length-converged intrinsic values.}
    \label{fig:placeholder}
\end{figure}

Because the central region has a finite channel length, the anharmonic NEGF calculation directly yields the two-terminal thermal conductance \(G(T,L,d)\) of a finite device. The quantity reported in the main text is therefore the effective finite-length thermal conductivity,
\[
\kappa_{\mathrm{eff}}(T,L,d)=\frac{G(T,L,d)L}{A(d)},
\]
where \(A(d)\) is the effective cross-sectional area. To assess the role of channel length, fully anharmonic NEGF calculations were performed for the H-passivated [001]-oriented Si nanowire with \(d=6.24\) nm at 300 K for \(L=\) 10, 12, 15, and 20 nm. As shown in Fig. S3, \(\kappa_{\mathrm{eff}}\) increases with channel length, demonstrating that transport remains length-dependent over the investigated range. Therefore, the results reported for \(L=10\) nm in the main text should be interpreted as finite-length effective thermal conductivities rather than length-converged intrinsic values. A systematic length-convergence study over the complete diameter range is computationally prohibitive within the fully anharmonic NEGF formalism.

\section{Numerical Uncertainty of Harmonic NEGF Calculations}

\begin{figure}[!htb]
    \centering
    \includegraphics[width=0.5\linewidth]{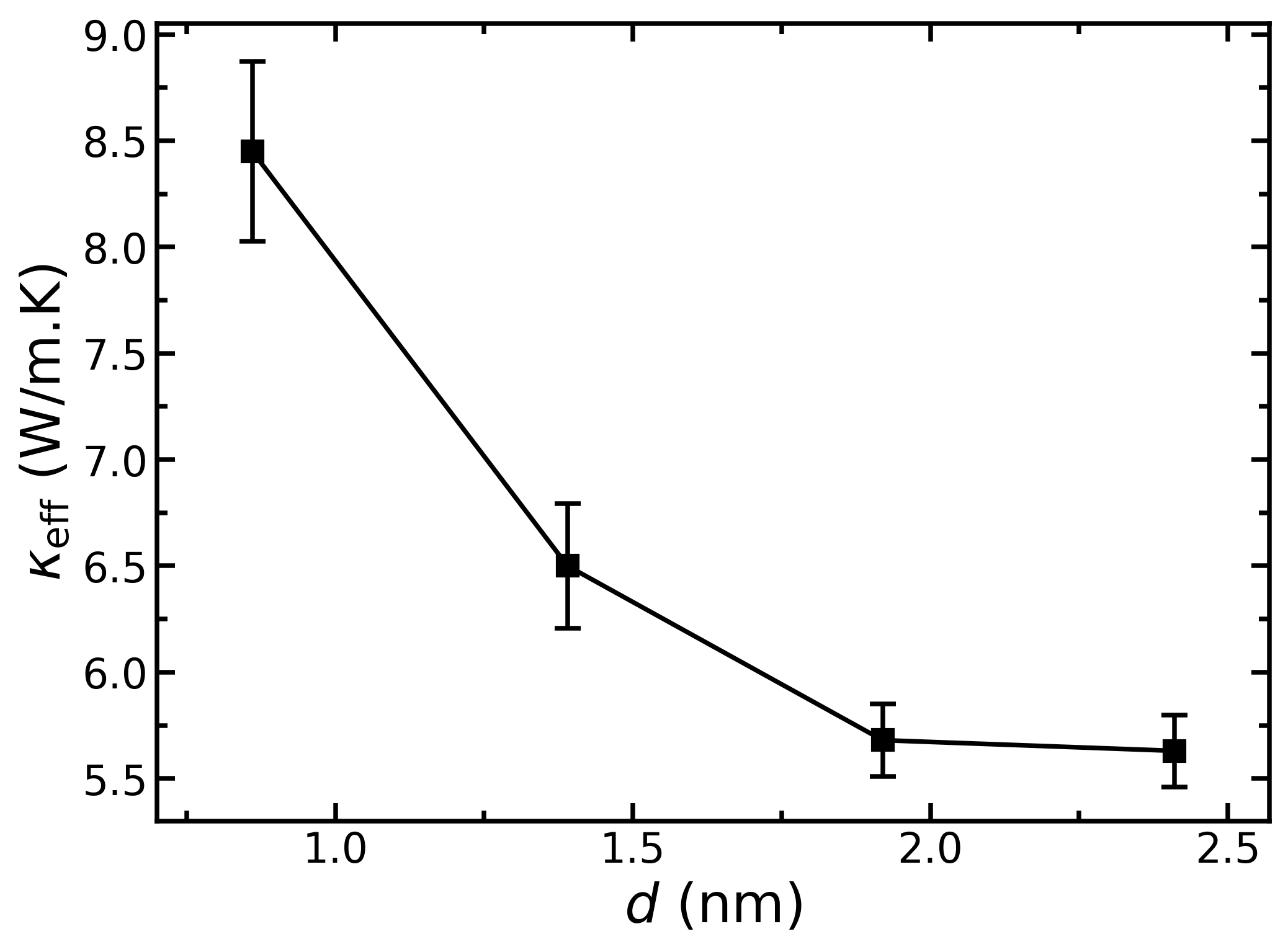}
    \caption{Numerical uncertainty of the effective thermal conductivity for H-passivated [001]-oriented Si nanowires with diameters \(d=0.86\), 1.39, 1.92, and 2.41 nm. Symbols show the mean values from five harmonic NEGF calculations using different converged combinations of Brillouin-zone \(k\)-point sampling and numerical broadening parameter \(\sigma\). Error bars indicate the spread among these calculations.}
    \label{fig:placeholder}
\end{figure}

To assess numerical uncertainty associated with the transport discretization, harmonic NEGF calculations were performed for H-passivated [001]-oriented Si nanowires with \(d=0.86\), 1.39, 1.92, and 2.41 nm. For each diameter, we performed five calculations using converged combinations of Brillouin-zone \(k\)-point sampling and the numerical broadening parameter \(\sigma\). The spread among the resulting values of \(\kappa_{\mathrm{eff}}\) provides an estimate of the numerical uncertainty associated with these parameters.

The effective thermal conductivities are \(8.45\pm0.42\), \(6.50\pm0.29\), \(5.68\pm0.17\), and \(5.63\pm0.17\) W m\(^{-1}\) K\(^{-1}\) for \(d=0.86\), 1.39, 1.92, and 2.41 nm, respectively. The uncertainty decreases with diameter over this range because the denser phonon spectrum reduces the integrated conductance's sensitivity to numerical discretization. This analysis quantifies numerical convergence of the harmonic NEGF framework with respect to \(k\)-point sampling and \(\sigma\); it does not include uncertainty associated with the anharmonic approximation or the interatomic potential.
\section{Temperature-Dependent Cryogenic Spectral Conductance}

\begin{figure}[!htb]
    \centering
    \includegraphics[width=\linewidth]{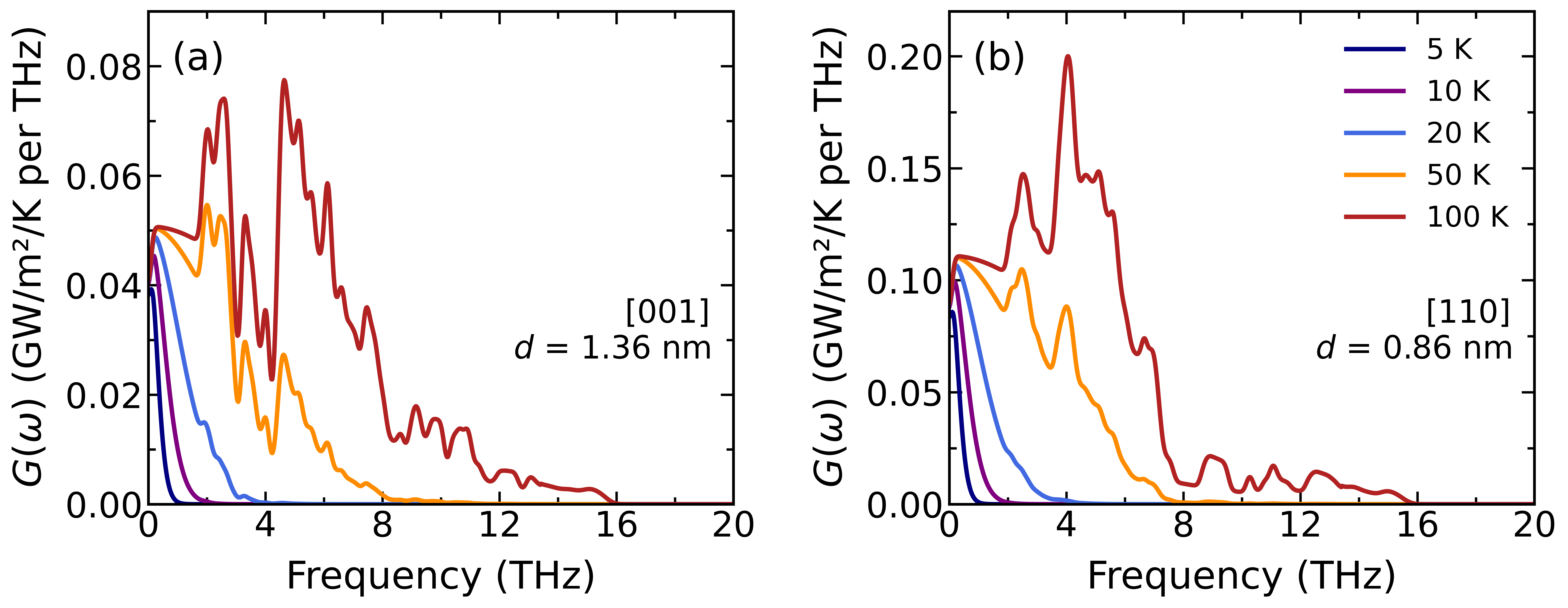}
    \caption{Temperature-dependent anharmonic NEGF spectral thermal conductance, \(G(\omega)\), for representative ultrathin H-passivated Si nanowires: (a) [001]-oriented nanowire with \(d=1.36\) nm and (b) [110]-oriented nanowire with \(d=0.86\) nm. Curves correspond to \(T=5\), 10, 20, 50, and 100 K. At 5–10 K, the thermally active spectral window is restricted primarily to the lowest-frequency confined acoustic-like modes. As temperature increases, higher-frequency confined subbands progressively contribute to the conductance.}
    \label{fig:placeholder}
\end{figure}

To further examine the cryogenic transport regime, the spectral thermal conductance \(G(\omega)\) was calculated at \(T=5\), 10, 20, 50, and 100 K for representative ultrathin H-passivated Si nanowires: a [001]-oriented nanowire with \(d=1.36\) nm and a [110]-oriented nanowire with \(d=0.86\) nm. As shown in Fig. S5, at 5–10 K the thermally active spectrum is restricted primarily to the lowest-frequency confined acoustic-like modes, whereas Bose–Einstein occupation strongly suppresses high-frequency phonons. As temperature increases, the active spectral window broadens, and additional confined subbands progressively contribute to the conductance.

The pronounced narrowing of the thermally active spectrum in the ultrathin limit is consistent with a few-mode, quasi-one-dimensional transport character at cryogenic temperatures. The term “quasi-one-dimensional” is used because the nanowires retain finite cross sections and multiple confined phonon subbands. These results support the interpretation that low-temperature transport is governed primarily by quantum occupation, confinement-modified low-frequency modes, and finite-device transmission, rather than by thermally activated Umklapp scattering.
\section{Harmonic and Anharmonic Comparison at Low Temperatures}

To distinguish the harmonic quantum-statistical contribution from the anharmonic correction at low temperature, harmonic Landauer and fully anharmonic NEGF spectral thermal conductances were compared for the representative [001]-oriented nanowire with \(d=1.36\) nm and [110]-oriented nanowire with \(d=0.86\) nm. As shown in Fig. S6, the harmonic and anharmonic spectra at 10 K occupy the same low-frequency confined acoustic-like transport window. Thus, quantum occupation and the confined harmonic phonon spectrum establish the primary cryogenic heat-carrying channels.

\begin{figure}[!t]
    \centering
    \includegraphics[width=\linewidth]{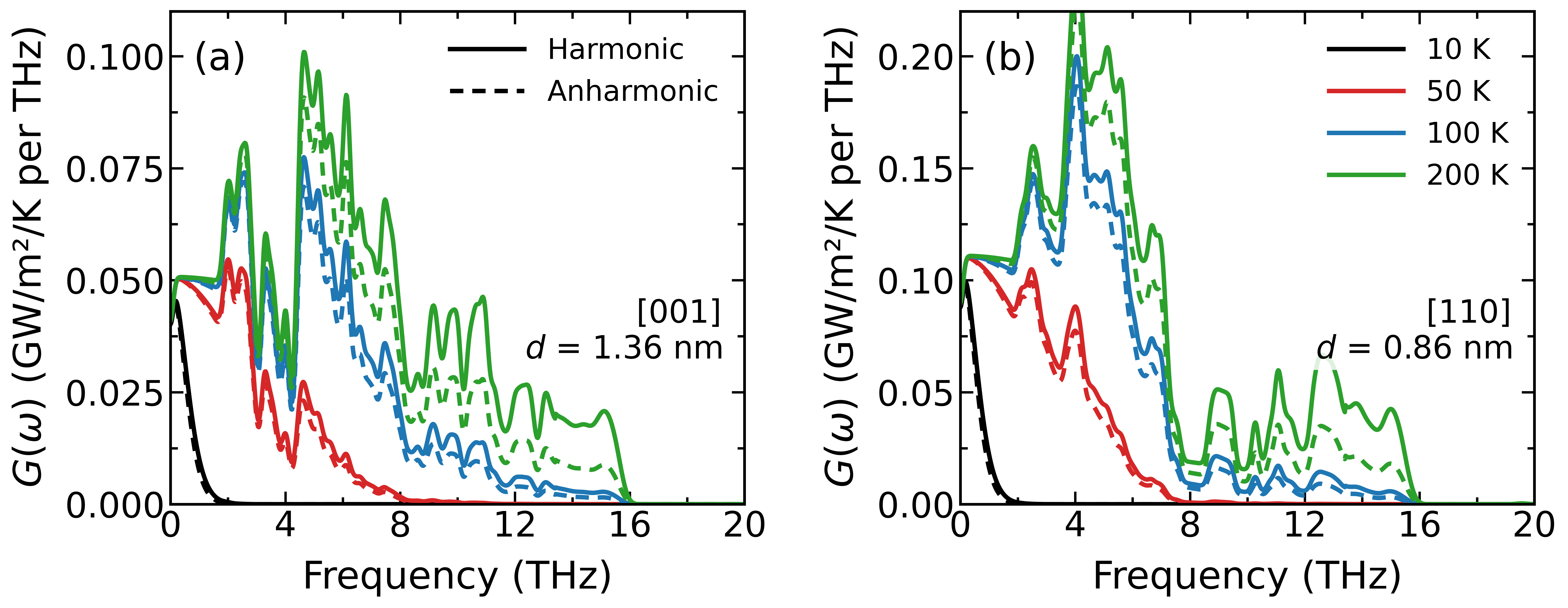}
    \caption{Comparison of harmonic Landauer and fully anharmonic NEGF spectral thermal conductances, \(G(\omega)\), for representative ultrathin H-passivated Si nanowires: (a) [001]-oriented nanowire with \(d=1.36\) nm and (b) [110]-oriented nanowire with \(d=0.86\) nm. Solid and dashed curves denote harmonic and anharmonic results, respectively. Black, red, and blue curves correspond to 10, 50, 100, and 200 K, respectively. At 10 K, low-frequency confined acoustic-like modes dominate heat transport. Anharmonicity produces a finite suppression of the spectral conductance relative to the harmonic quantum Landauer result.}
    \label{fig:placeholder}
\end{figure}

\begin{table}[!hb]
\centering
\caption{Harmonic Landauer and fully anharmonic NEGF effective thermal
conductivities, $\kappa_{\mathrm{eff}}$, and the corresponding reduction
due to anharmonicity for representative H-passivated Si nanowires with
channel length $L=10$ nm.}
\label{tab:harmonic_anharmonic_kappa}

\begin{ruledtabular}
\begin{tabular}{cccccc}
Orientation & $d$ (nm) & $T$ (K) &
\multicolumn{1}{c}{Harmonic $\kappa_{\mathrm{eff}}$} &
\multicolumn{1}{c}{Anharmonic $\kappa_{\mathrm{eff}}$} &
\multicolumn{1}{c}{Reduction} \\
& & & \multicolumn{2}{c}{(W/m$\cdot$K)} & (\%) \\
\hline
$\left[001\right]$ & 1.36 & 10  & 0.34 & 0.32 & 5.9 \\
$\left[001\right]$ & 1.36 & 50  & 2.07 & 1.92 & 7.2 \\
$\left[001\right]$ & 1.36 & 100 & 4.47 & 4.02 & 10.1 \\
$\left[001\right]$ & 1.36 & 200 & 6.99 & 5.66 & 19.0 \\
\hline
$\left[110\right]$ & 0.86 & 10  & 0.75 & 0.72 & 4.0 \\
$\left[110\right]$ & 0.86 & 50  & 4.75 & 4.51 & 5.1 \\
$\left[110\right]$ & 0.86 & 100 & 9.43 & 8.65 & 8.3 \\
$\left[110\right]$ & 0.86 & 200 & 13.67 & 11.45 & 16.2 \\
\end{tabular}
\end{ruledtabular}
\end{table}

Anharmonic phonon–phonon interactions produce a finite reduction in \(\kappa_{\mathrm{eff}}\) relative to the harmonic Landauer result, as summarized in Table S1. The anharmonic correction remains finite at 50, 100, and 200 K as additional confined modes become thermally active. These calculations therefore separate the dominant harmonic quantum baseline from the additional anharmonic correction and show that the low-temperature diameter dependence should not be attributed to strong thermally activated Umklapp scattering.



\bibliography{references.bib}